\documentclass{cimento1}

\usepackage{graphicx}  % got figures? uncomment this

\newcommand{\bea}{\begin{eqnarray}}
\newcommand{\eea}{\end{eqnarray}}
\newcommand{\beq}{\begin{equation}}
\newcommand{\eeq}{\end{equation}}
\newcommand{\nn}{\nonumber}
\newcommand{\gev}{{\rm GeV}}
\newcommand{\mev}{{\rm MeV}}
\newcommand{\msb}{\overline{\rm{MS}}}

\def\simge{\mathrel{\rlap{\raise 0.511ex \hbox{$>$}}{\lower 0.511ex
 \hbox{$\sim$}}}}
\def\simle{\mathrel{\rlap{\raise 0.511ex \hbox{$<$}}{\lower 0.511ex
 \hbox{$\sim$}}}}
\def\slash#1{\setbox0=\hbox{$#1$}\dimen0=\wd0 \setbox1=\hbox{/} \dimen1=\wd1
 \ifdim\dimen0>\dimen1 \rlap{\hbox to \dimen0{\hfil/\hfil}} #1
 \else \rlap{\hbox to \dimen1{\hfil$#1$\hfil}} / \fi}

\title{Flavour physics and Lattice QCD: averages of lattice inputs for the
Unitarity Triangle Analysis}
\shorttitle{Flavour physics and Lattice QCD: averages of lattice inputs for the
UTA}
\author{V.~Lubicz\from{roma3}% \ETC
\thanks{Speaker}
\atque
C.~Tarantino\from{roma3}}
\instlist{\inst{roma3} Dip. di Fisica, Universit\`a di Roma
Tre, Via della Vasca Navale 84, I-00146 Rome, Italy.}
\PACSes{\PACSit{12.38.Gc}{Lattice QCD calculations}}
\begin{document}

\maketitle

\begin{abstract}
We review recent results of Lattice QCD calculations relevant for flavour
physics. We discuss in particular the hadronic parameters entering the
amplitudes of $K^0-\bar K^0$, $D^0-\bar D^0$ and $B^0-\bar B^0$ mixing, the
$B$- and $D$-meson decay constants and the form factors controlling $B$-meson
semileptonic decays. On the basis of these lattice results, which are
extensively collected in the paper, we also derive our averages of the relevant
hadronic parameters.
\end{abstract}

The aim of this talk is to illustrate the status of the art of lattice QCD
calculations relevant for flavour physics and to provide averages of the
hadronic quantities which are useful for phenomenological analysis of flavour
physics, within or beyond the Standard Model. These averages will be used in
particular by the ${\rm UT}fit$ collaboration in the unitarity triangle analysis
(UTA)~\cite{UTfit-ifae}-\cite{wwwutfit}. We believe, however, that the whole
discussion given in the present contribution can be of more general interest.
Even though the process of averaging lattice results unavoidably involves some
degree of subjectivity, for all the quantities discussed in this work we also
provide extensive collections of lattice results, as well as the references to
the original papers. In addition, we try to enrich the discussion with some
technical considerations about the accuracy of the various lattice calculations,
which can be useful for the non-expert of the field in order to have a deeper
view and to draw their conclusions.

The status of the art of Lattice QCD calculations is represented by unquenched
si\-mu\-la\-tions performed with either $N_f=2$ or $N_f=2+1$ dynamical quarks
and values of the light (i.e. up/down) quark masses well below $m_s/2$, where
$m_s$ is the physical strange quark mass. Both these features are particularly
important to keep the lattice systematic uncertainties under control. On the one
hand, it is well known that the quenched ap\-pro\-xi\-ma\-tion introduces a
systematic error which, besides depending on the specific quantity under
investigation, is also difficult to evaluate (if not by performing the
corresponding unquenched calculation). On the other hand, lattice simulations
performed with sufficiently light values of the up and down quark masses
($m_{ud} < m_s/2$) allow to rely on the predictions of chiral perturbation
theory when performing the chiral extrapolations of the lattice results to the
physical masses.

If the possibility of performing unquenched calculations at light quark masses
represents a major advance of Lattice QCD in the last few years, it has also to
be noted that these calculations have not always reached yet the same accuracy
in controlling other sources of systematic effects as done in the more recent
(and less expensive) quenched calculations. For instance, some of the unquenched
results have not yet been extrapolated to the continuum limit, and
non-perturbative renormalization techniques have not always been implemented.
Moreover, while most of the quenched results have been derived by many lattice
collaborations, often using different approaches, in several cases only few (if
any) unquenched results are already available for a given quantity. For
instance, the unquenched results for the kaon B-parameter have been only
obtained so far at fixed (and rather large) values of the lattice spacing,
whereas the matrix elements of the full basis of four-fermion operator relevant
for $K^0-\bar K^0$ mixing in generic New Physics models have been only obtained
so far within the quenched approximation.

In what follows we illustrate in some detail the results for various lattice
parameters and present our averages. We will not be able to discuss in the
present contribution the whole set of hadronic parameters available from the
lattice and which are relevant for flavour physics. In particular, we are not
going to review the accurate lattice results for the quark masses, for the ratio
of decay constants $f_K/f_\pi$ and for the vector form factor controlling the
semileptonic $K\ell 3$ decays. For the latter two quantities, which allow the
important determination of $V_{us}$ (the Cabibbo angle), we refer to the review
by A.~Juttner at the Lattice'07 conference~\cite{Juttner:2007sn}.

\section{The kaon $B$-parameter $B_K$}
\begin{figure}[t]
\begin{center}
\hspace{-0.6cm}
\includegraphics[scale=0.26,angle=270]{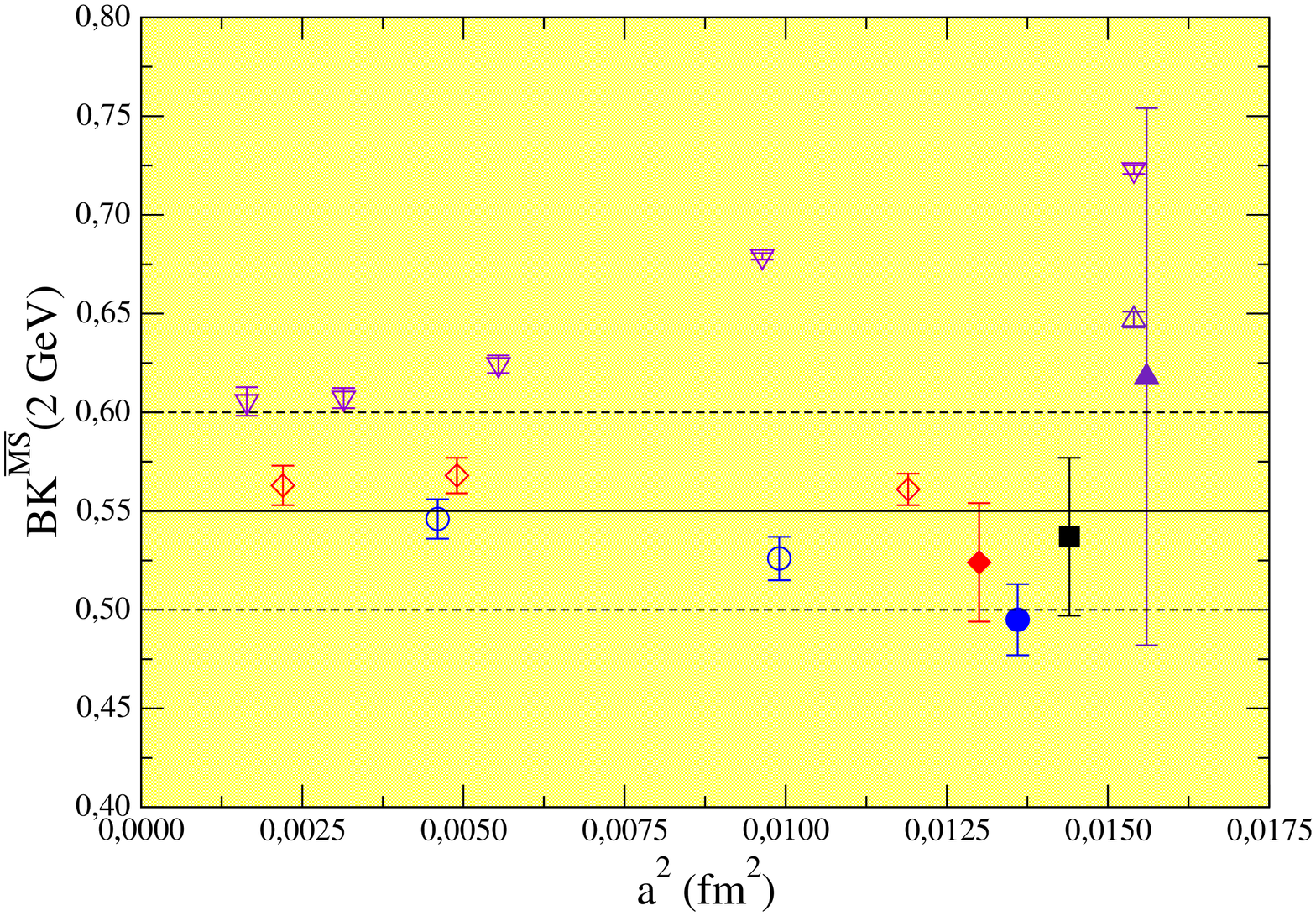}
\hspace{-0.8cm}
\includegraphics[scale=0.26,angle=270]{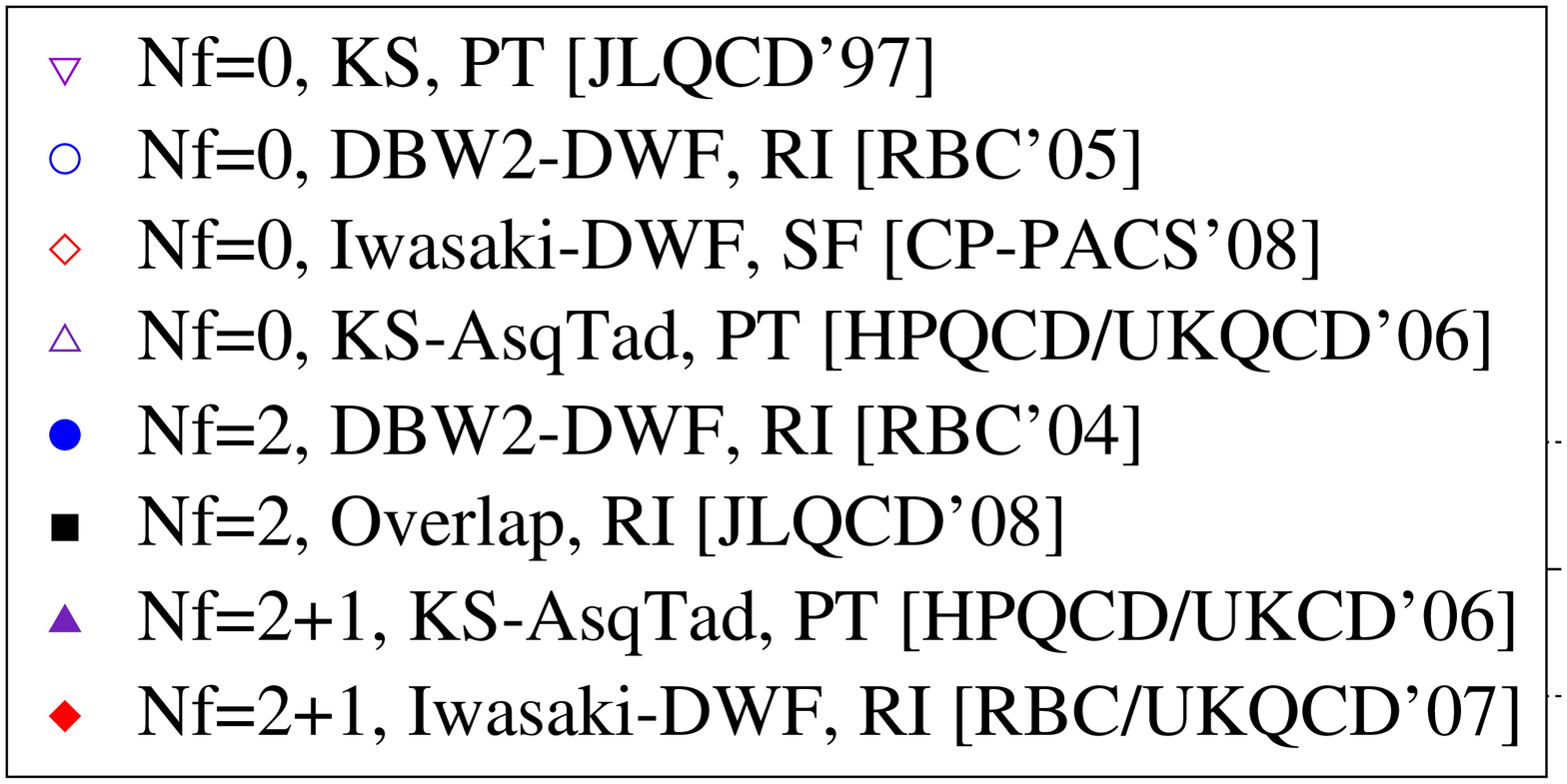}
\end{center}
\vspace{-0.5cm}
\caption{{\sl Lattice QCD results for $B_K^{\msb}(2\
\gev)$~\cite{Aoki:1997nr}-\cite{Antonio:2007pb} plotted as a function of the
lattice spacing square. All results selected in the plot are ${\cal
O}(a)$-improved and required only multiplicative renormalization of the relevant
$\Delta S=2$ four-fermion operator. Filled points are unquenched results
obtained with either $N_f=2$ or $N_f=2+1$ dynamical quarks. Empty points are
quenched results. Our average is shown by the horizontal band.}}
\label{fig:BK}
\end{figure}

A collection of quenched and unquenched lattice results for $B_K^{\msb}(2\
\gev)$~\cite{Aoki:1997nr}-\cite{Antonio:2007pb} is shown in fig.\ref{fig:BK} as
a function of the square of the lattice spacing $a$. Among the quenched results
we have selected all determinations which are ${\cal O}(a)$-improved, i.e. that
are affected by discretization errors of ${\cal O}(a^2)$ or smaller, and have
been obtained at several values of the lattice spacing, thus allowing an
extrapolation to the physical continuum limit. An additional feature of all
results shown in fig.\ref{fig:BK} is that they have been obtained with lattice
discretizations of the fermionic action such that the $\Delta S=2$ four-fermion
operator defining $B_K$ is subject to multiplicative renormalization only. This
allows to significantly increase the accuracy of the determination of the kaon
$B$-parameter.
%%%%%%%%%%%%%%%%%%%%%%%%%%%%%%%%%%%%%%%%%%%%%%%%%%%%%%%%%%%%%%%%%%
%%%%%%%%%%%%%%%%%%%%%%%%%%%%%%%%%%%%%%%%%%%%%%%%%%%%%%%%%%%%%%%%%%

As illustrated in fig.\ref{fig:BK}, the unquenched results for $B_K$,
represented by filled points in the plot, have been all obtained so far at a
fixed value of the lattice spacing. Thus, at variance with the quenched
determinations, the unquenched results cannot be extrapolated to the continuum
limit, and the size of discretization errors affecting them cannot be easily
estimated. On general grounds, we note that there is no reason to expect
discretization errors in the unquenched case to have smaller size than those
affecting the quenched results.

The comparison between quenched and unquenched determinations of $B_K$ obtained
with the same lattice action and at similar values of the lattice spacing does
not show any evidence of the quenching effect within the present uncertainties.
Moreover, the comparison between $N_f=0$ and $N_f=2$ results obtained with the
same lattice action but at different values of the lattice spacing, see for
instance the case of the DBW2-DWF action (empty and filled circles in the plot),
suggests that quenched and unquenched results could be affected by similar
discretization errors. This observation has motivated the RBC collaboration to
consider their $N_f=0$ and $N_f=2$ results together in order to perform a
combined extrapolation to the continuum limit~\cite{Aoki:2005ga}. By taking into
account the scale dependence suggested by the quenched results, we quote as
final average for $B_K$ the value
\beq
\label{eq:bkms}
B_K^{\msb}(2\ \gev) = 0.55 \pm 0.05 \ ,
\eeq
which corresponds to the renormalization group invariant parameter
\beq
\widehat B_K = 0.75 \pm 0.07 \ .
\eeq
Clearly, a better estimate of discretization errors in the unquenched case will
be obtained once the results of unquenched simulations performed at different
values of the lattice spacing, which are currently in progress, will be
available.

Our average in eq.~(\ref{eq:bkms}) is shown by the horizontal band in
fig.\ref{fig:BK}. It can be compared with the average $B_K^{\msb}(2\ \gev) =
0.58 \pm 0.03 \pm 0.06$ quoted by Dawson at Lattice'05~\cite{dawson}, where the
central value and the first error corresponded to the average of the quenched
results in the continuum limit, while the second error was an estimate of the
quenching effect. The recent unquenched results for $B_K$ allow to remove the
latter uncertainty and slightly decrease the previous central value.

\section{The $B$-mesons decay constants $f_{B_s}$ and $f_B$}
\begin{figure}[t]
\begin{center}
\hspace{-0.6cm}
\includegraphics[scale=0.26,angle=270]{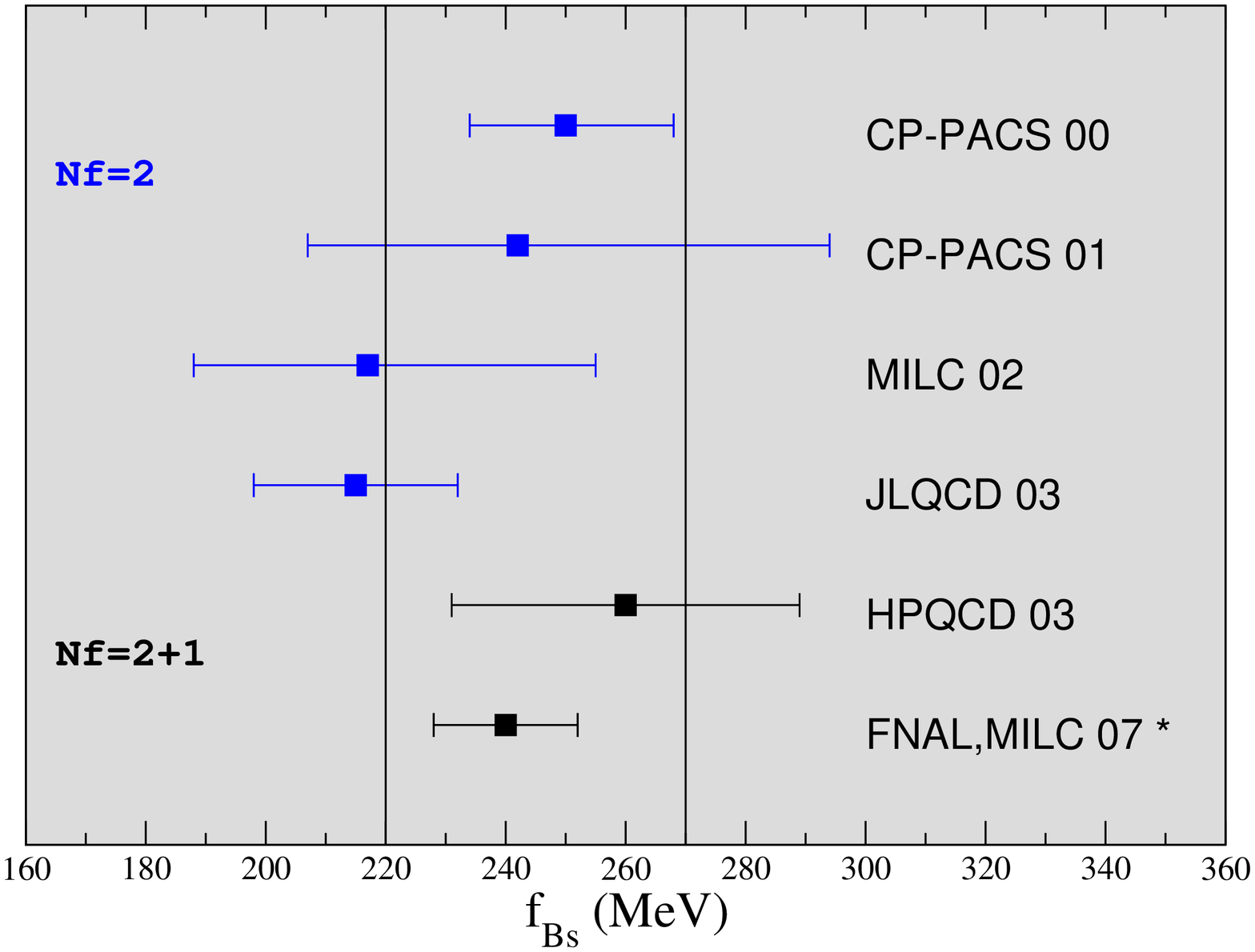}
\hspace{-0.8cm}
\includegraphics[scale=0.26,angle=270]{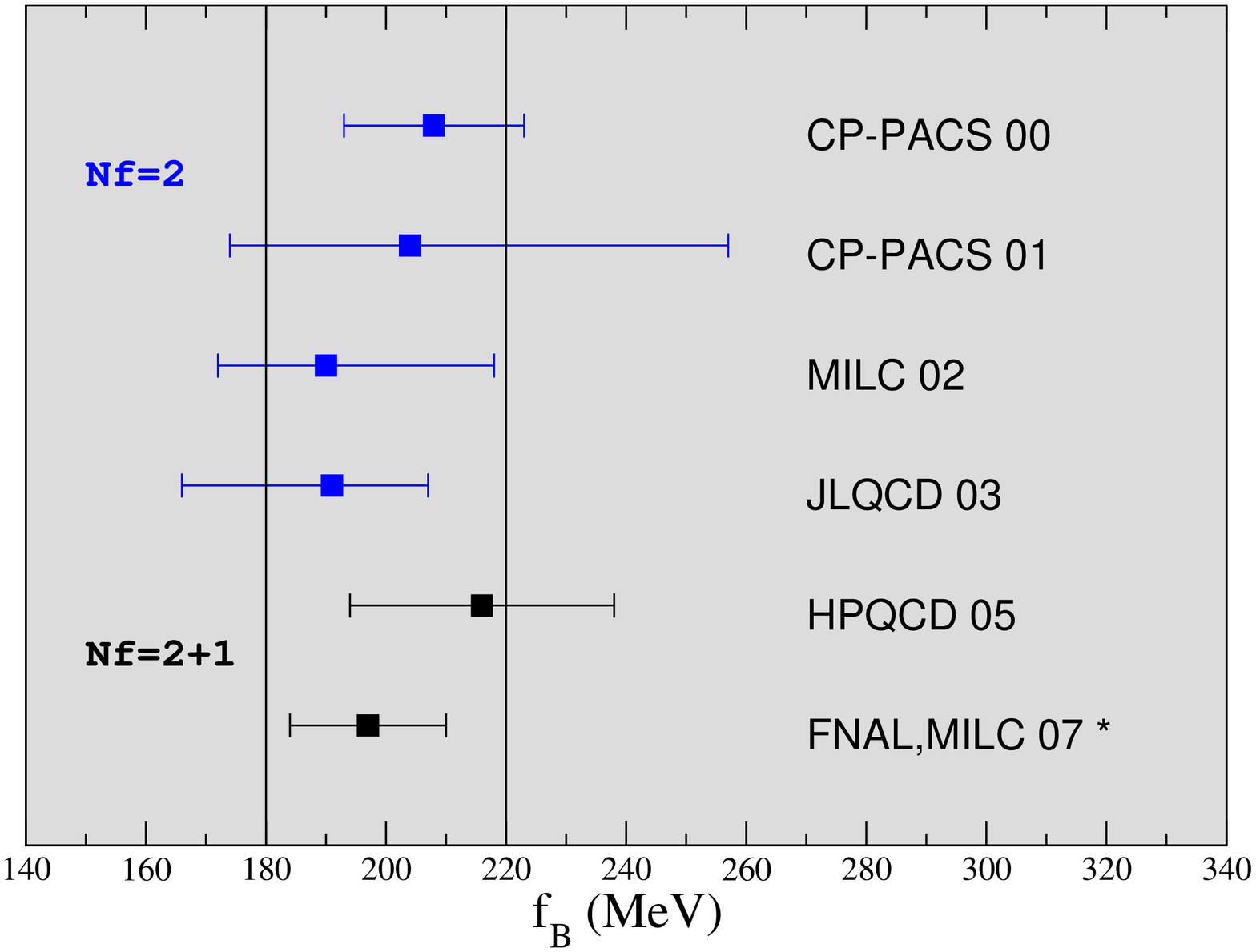}\\
\vspace{-0.3cm}
\includegraphics[scale=0.26,angle=270]{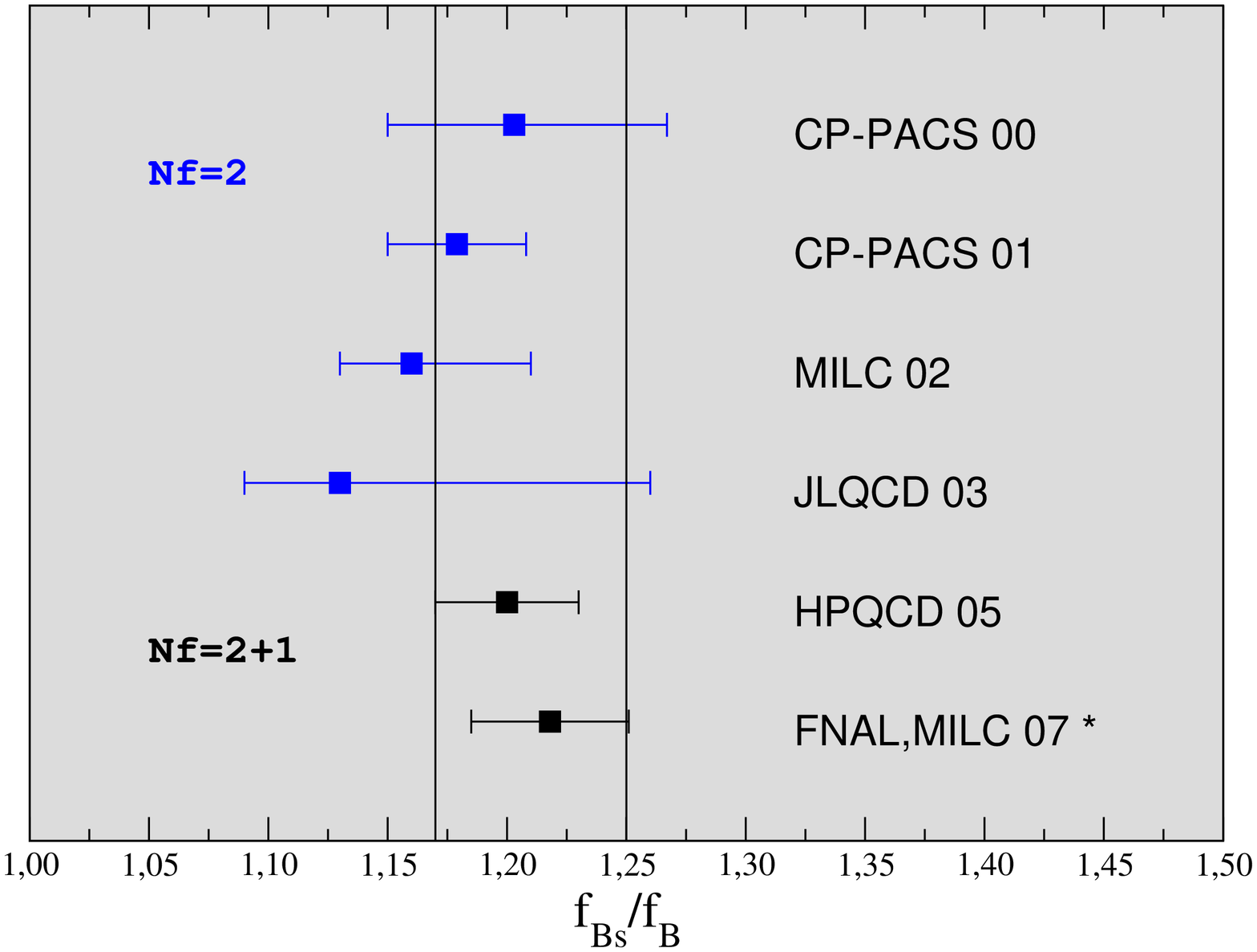}
\end{center}
\vspace{-0.5cm}
\caption{\sl Lattice QCD results for the $B$-mesons decay constants $f_{B_s}$
(top left), $f_{B}$ (top right) and the ratio $f_{B_s}/f_{B}$ (bottom) obtained
with $N_f=2$ and $N_f=2+1$ unquenched
simulations~\cite{AliKhan:2000eg}-\cite{Bernard:2007}. A star in the legends
labels preliminary results. Our averages are shown by the vertical bands.}
\label{fig:fbssfb}
\end{figure}

The unquenched lattice QCD results for the $B$-mesons decay constants $f_{B_s}$
and $f_{B}$ and for their ratio
$f_{B_s}/f_{B}$~\cite{AliKhan:2000eg}-\cite{Bernard:2007} are shown in
fig.\ref{fig:fbssfb}. They have been obtained by treating the $b$ quark on the
lattice with two different approaches, either FNAL~\cite{ElKhadra:1996mp} or
NRQCD.
%%%%%%%%%%%%%%%%%%%%%%%%%%%%%%%%%%%%%%%%%%%%%%%%%%%%%%%%%%%%%%%%%%%%%
%%%%%%%%%%%%%%%%%%%%%%%%%%%%%%%%%%%%%%%%%%%%%%%%%%%%%%%%%%%%%%%%%%%%%

Our average for the $B_s$ decay constant,
\beq
\label{eq:fbsave}
f_{B_s}=245 \pm 25 \ \mev\ ,
\eeq
is based on the results shown in the upper plot of fig.\ref{fig:fbssfb}. The
lattice determination of $f_B$ is more delicate, because its value is enhanced
by chiral logs effects relevant at low quark masses. In order to properly
account for these effects, simulations at light values of the quark mass
(typically $m_{ud}<m_s/2$) are required. For this reason, we derive our average
for $f_{B}$ and for the ratio $f_{B_s}/f_{B}$ by taking into account only the
results HPQCD'05~\cite{Gray:2005ad} and FNAL/MILC'07~\cite{Bernard:2007} (see
fig.\ref{fig:fbssfb}), which use the MILC gauge field configurations generated
at light quark masses as low as $m_s/8$. In this way we obtain
\beq
f_{B_s}/f_B = 1.21 \pm 0.04 \ .
\eeq
Finally, we derive directly from the previous averages of $f_{B_s}$ and
$f_{B_s}/f_{B}$ the estimate of the $B$-meson decay constant
\beq
\label{eq:fbave}
f_B=200 \pm 20 \ \mev \ .
\eeq

\section{The $D$-mesons decay constants $f_{D_s}$ and $f_D$}
The CKM matrix elements $V_{cs}$ and $V_{cd}$, which control the rate of
leptonic $D_s^+ \to \ell^+ \nu_\ell$ and $D^+ \to \ell^+ \nu_\ell$ decays, are
well constrained by the unitarity of the CKM matrix, which predicts $|V_{cs}|
\simeq 1-\lambda^2/2$ and $|V_{cd}| \simeq \lambda$ (up to ${\cal O}(\lambda^4)$
corrections), where $\lambda$ is the Wolfenstein parameter. By using these
constraints, the measurements of the leptonic decay rates then allow a
determination of the $D$-mesons decay constants, $f_{Ds}$ and $f_{D}$. The
current experimental averages, as evaluated in ref.~\cite{Rosner:2008yu}, are
\bea
\label{eq:fdsexp}
&& f_{D_s}^{EXP.}=273 \pm 10 \ \mev \quad , \quad 
f_D^{EXP.}=205.8 \pm 8.9 \ \mev \quad , \\
&& \qquad \qquad \qquad 
\left(f_{D_s}/f_D\right)^{EXP.} = 1.33 \pm 0.07 \ .\nn 
\eea

\begin{figure}[t]
\begin{center}
\includegraphics[scale=0.30,angle=270]{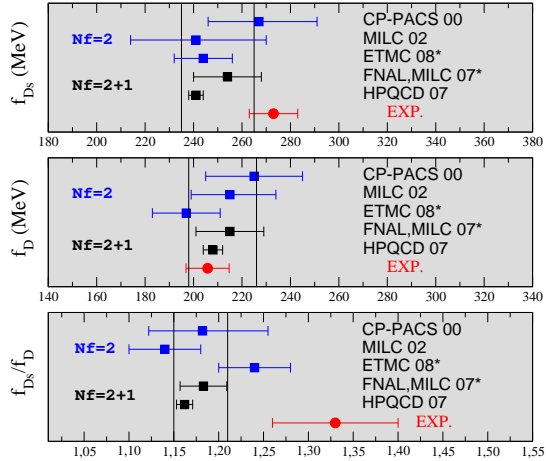}
\end{center}
\vspace{-0.5cm}
\caption{\sl Lattice QCD results for the $D$-mesons decay constants $f_{D_s}$
(top), $f_{D}$ (center) and for the ratio $f_{D_s}/f_{D}$ (bottom) obtained
with $N_f=2$ and $N_f=2+1$ unquenched
simulations~\cite{AliKhan:2000eg,Bernard:2002pc,Bernard:2007, cecilia_lat08,
Follana:2007uv}. A star in the legends labels preliminary results. Our averages
are shown by the vertical bands. The corresponding experimental
averages~\cite{Rosner:2008yu} are also shown for comparison.}
\label{fig:fdsfd}
\end{figure}

The unquenched lattice QCD results for these
constants~\cite{AliKhan:2000eg,Bernard:2002pc,Bernard:2007,cecilia_lat08,
Follana:2007uv} are shown in fig.\ref{fig:fdsfd}, where they are also compared
with the experimental averages given in eq.~(\ref{eq:fdsexp}).
%%%%%%%%%%%%%%%%%%%%%%%%%%%%%%%%%%%%%%%%%%%%%%%%%%%%%%%%%%%%%%%%%%%%%
%%%%%%%%%%%%%%%%%%%%%%%%%%%%%%%%%%%%%%%%%%%%%%%%%%%%%%%%%%%%%%%%%%%%%

Among the various lattice determinations, which show overall a very good
consistency among each other, it deserves to be noted the result of the HPQCD
collaboration~\cite{Follana:2007uv}, which quotes $f_{D_s}$ and $f_{D}$ with
the impressive accuracy of about 1\% and 2\% respectively. Remarkably, while the
result for $f_{D}$ of ref.~\cite{Follana:2007uv} is in good agreement with the
experimental average, the prediction for $f_{D_s}$ differs from the experimental
value by approximately 3 standard deviations. This discrepancy has been
interpreted in~\cite{Dobrescu:2008er} as an evidence for New Physics effects in
leptonic $D_s$ decay.

A critical review of the lattice study of ref.~\cite{Follana:2007uv} is not
possible yet, since important details on the analysis are not given in the
paper, being postponed to a longer publication~\cite{Follana:2007uv}. Of
particular relevance, in this respect, are the details of the Bayesian fits
implemented to perform the combined continuum and chiral extrapolation, as well
as the knowledge of the precise functional form assumed for this extrapolation.
The large number of free parameters introduced in the fit (specifically, 45
parameters with 28 data points) is the source of some concern, particularly in
view of the remarkable accuracy claimed on the final results. We also mention
that the use of the so called ``fourth root trick" in dynamical simulations of
staggered quarks, which the results of ref.~\cite{Follana:2007uv} are based on,
is controversial. This approach has been the subject of intensive theoretical
and numerical investigation in the last few years. Despite the valuable
progresses achieved in our understanding of the rooting procedure, the dispute
on its validity has not been, and will likely never be, completely resolved (for
extensive discussions, see e.g. the recent
reviews~\cite{Sharpe:2006re,Kronfeld:2007ek} at the Lattice conferences).

Waiting for a confirmation of the precise predictions of
ref.~\cite{Follana:2007uv} by other lattice calculations, in deriving averages
for the $D$-mesons decay constants we choose to quote uncertainties which are
larger than those given in ref.~\cite{Follana:2007uv}, and comparable in size
to those presented by the other results. Thus, we quote as our final averages
\beq
\label{eq:fdslatt}
f_{D_s}=250 \pm 15 \ \mev \quad , \quad f_D=212 \pm 14 \ \mev \quad , \quad
f_{D_s}/f_D = 1.18 \pm 0.03 \ ,
\eeq
which are the values also shown by the vertical bands in fig.\ref{fig:fdsfd}.

\section{Matrix elements for $B^0_{d/s}-\bar B^0_{d/s}$ mixing}
\begin{figure}[t]
\begin{center}
\hspace{-0.6cm}
\includegraphics[scale=0.26,angle=270]{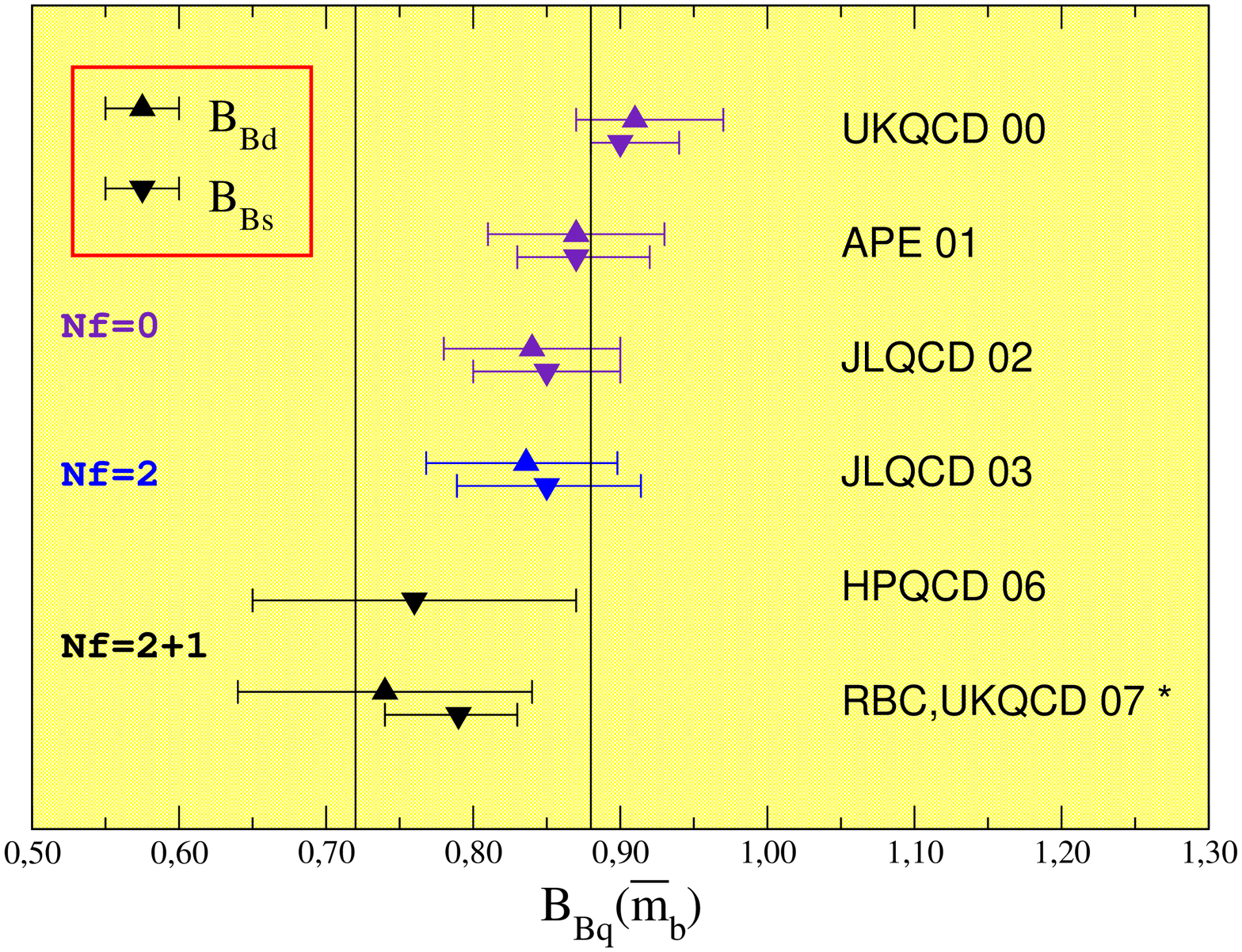}
\hspace{-0.8cm}
\includegraphics[scale=0.26,angle=270]{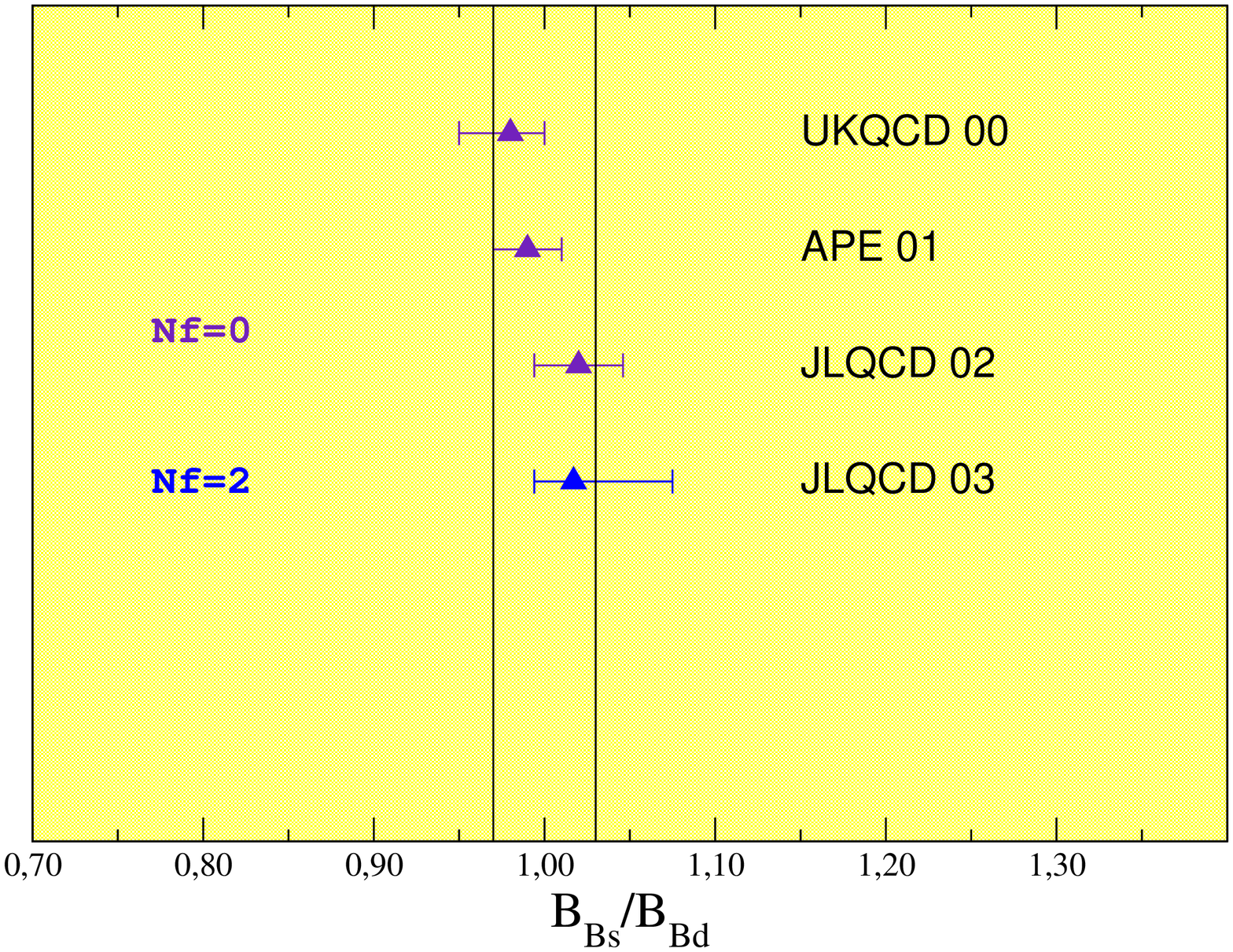}
\end{center}
\vspace{-0.5cm}
\caption{{\sl Lattice QCD results for $B_{B_d}^{\msb}(m_b)$ and
$B_{B_s}^{\msb}(m_b)$ (left) and for their ratio
$B_{B_s}/B_{B_d}$ (right)~\cite{Aoki:2003xb},
\cite{Lellouch:2000tw}-\cite{Albertus}. A star in the legend labels preliminary
results. Our averages are shown by the vertical bands.}}
\label{fig:bbds}
\end{figure}

A collection of quenched and unquenched lattice results for the $B_d$ and $B_s$
bag parameters and for their ratio~\cite{Aoki:2003xb},
\cite{Lellouch:2000tw}-\cite{Albertus} is shown in fig.\ref{fig:bbds}.
%%%%%%%%%%%%%%%%%%%%%%%%%%%%%%%%%%%%%%%%%%%%%%%%%%%%%%%%%%%%%%%%%%
%%%%%%%%%%%%%%%%%%%%%%%%%%%%%%%%%%%%%%%%%%%%%%%%%%%%%%%%%%%%%%%%%%
We firstly observe that the dependence on the light quark mass, that should
allow to distinguish between $B_d$ and $B_s$, is practically invisible.
Moreover, the unquenched results, obtained with $N_f=2$ and $N_f=2+1$ dynamical
quarks, tends to be slightly lower than the quenched determinations, though
still well compatible within the errors. On the basis of the unquenched results
we quote the average
\beq
\label{eq:bbms}
B_{B_d}^{\msb}(m_b) =  B_{B_s}^{\msb}(m_b) = 0.80 \pm 0.08 \ ,
\eeq
which is illustrated by the vertical band in fig.\ref{fig:bbds} (left) and
corresponds to the re\-nor\-ma\-li\-za\-tion group invariant parameters
\beq
\widehat B_{B_d} =  \widehat B_{B_s} = 1.22 \pm 0.12 \ .
\eeq
For the ratio $\widehat B_{B_s}/\widehat B_{B_d}$, where statistical and
systematic uncertainties partially cancel (see fig.\ref{fig:bbds} right), we
quote
\beq
\widehat B_{B_s}/\widehat B_{B_d} = 1.00 \pm 0.03 \ .
\eeq

Combining the above results with the averages quoted for the $B$-mesons
decay constants, eqs.~(\ref{eq:fbsave})-(\ref{eq:fbave}), we then obtain
\bea
\label{eq:bbarmix}
&& f_{B_s} \sqrt{\widehat B_{B_s}}=270 \pm 30 \ \mev \quad , \quad
f_{B} \sqrt{\widehat B_{B_d}}=225 \pm 25 \ \mev \quad , \nn \\
&& \qquad  \qquad \qquad
\xi=\frac{f_{B_s} \sqrt{\widehat B_{B_s}}}{f_{B} \sqrt{\widehat B_{B_d}}}=
1.21 \pm 0.04 \ .
\eea
It is important to compare the average of $f_{B_s} \sqrt{\widehat B_{B_s}}$ of
eq.~(\ref{eq:bbarmix}), obtained from $f_{B_s}$ and $B_{B_s}$ separately, with
the direct calculation of $f_{B_s} \sqrt{\widehat B_{B_s}}$ from the whole
$B_s$-mixing matrix element. For the latter, the unquenched results obtained in
JLQCD'03~\cite{Aoki:2003xb} and HPQCD'06~\cite{Dalgic:2006gp} are $f_{B_s}
\sqrt{\widehat B_{B_s}}=245 \pm 10 ^{+19}_{-17} \ \mev$ and $f_{B_s}
\sqrt{\widehat B_{B_s}}=281 \pm 21 \ \mev$ respectively, well consistent with
the average of eq.~(\ref{eq:bbarmix}).

\section{$V_{cb}$ exclusive}
\begin{figure}[t]
\begin{center}
%\vspace{-2.5cm}
\hspace{-0.5cm}
\includegraphics[width=5.0cm,height=7.0cm,angle=270]{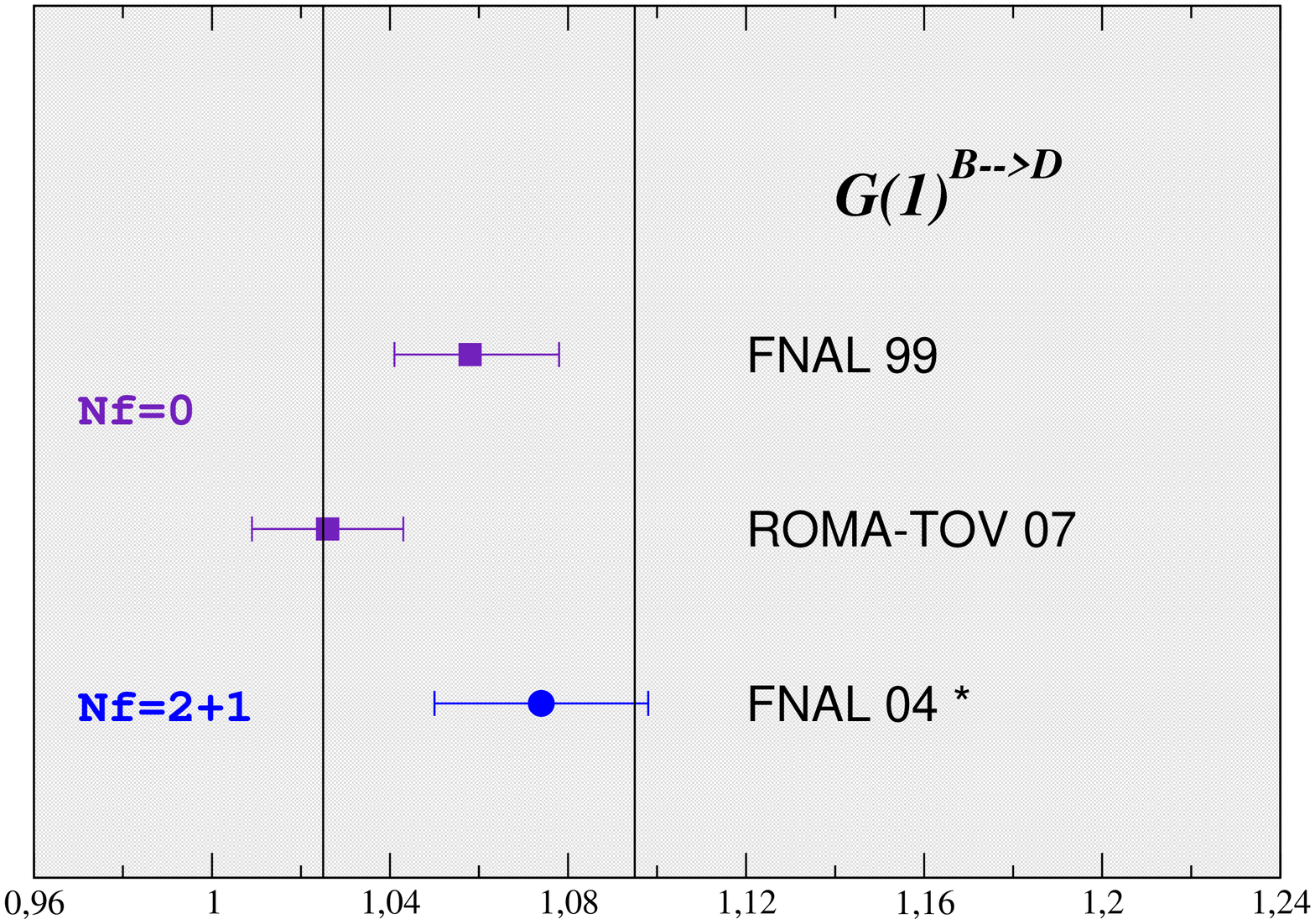}
\hspace{-0.5cm}
\includegraphics[width=5.0cm,height=7.0cm,angle=270]{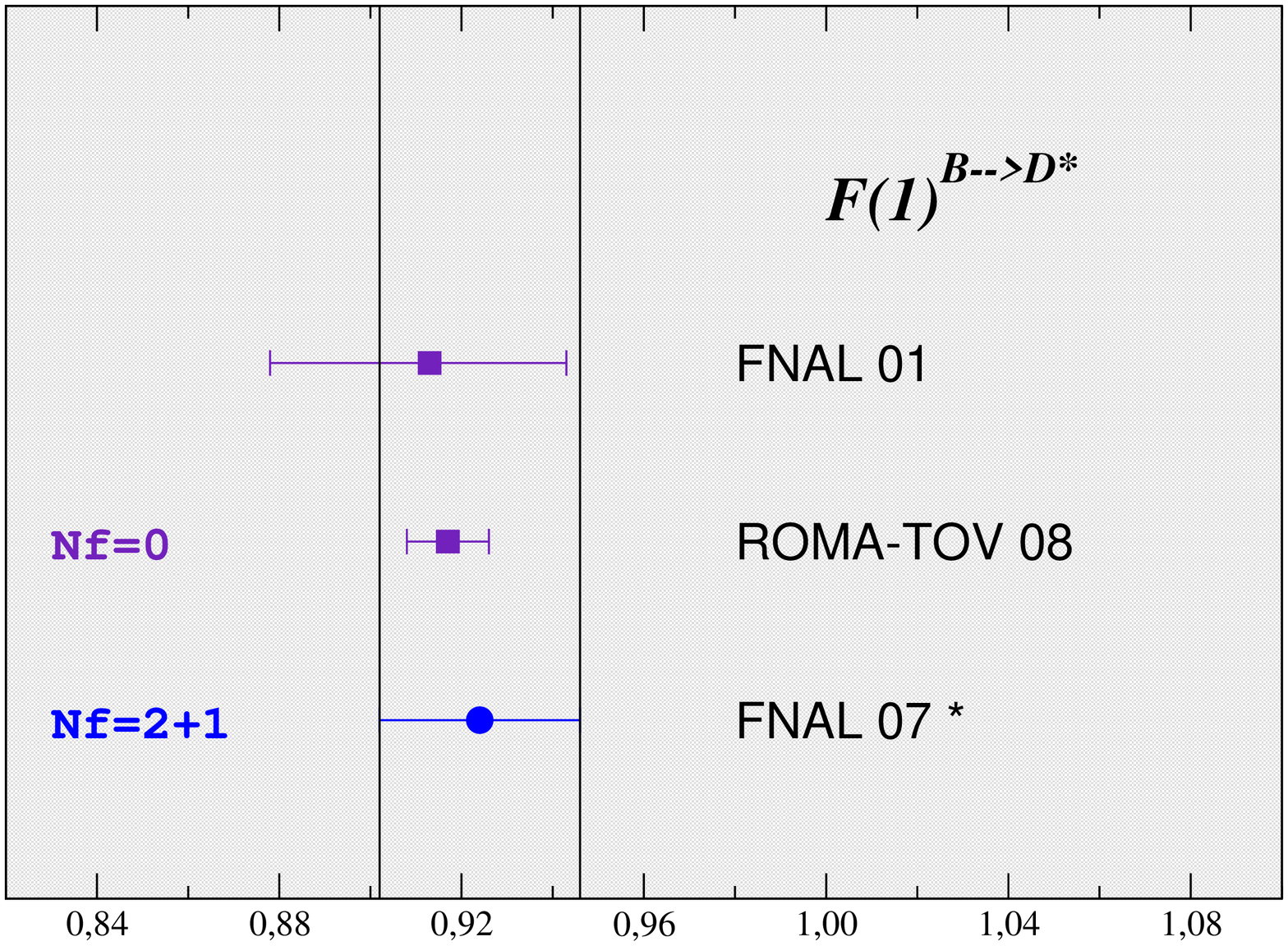}
\end{center}
\vspace{-0.5cm}
\caption{{\sl Lattice QCD results for the form factors $F(1)$ (left) and
$G(1)$ (right)~\cite{Hashimoto:2001nb}-\cite{Okamoto:2004xg} controlling $B\to
D^* l \nu$ and $B\to D l \nu$ decays. A star in the legend labels preliminary
results. Our averages are shown by the vertical bands.}}
\label{fig:Vcb}
\end{figure}

The lattice results for the zero momentum transfer form factors of $B\to D^* l
\nu$ and $B\to D l \nu$ semileptonic
decays~\cite{Hashimoto:2001nb}-\cite{Okamoto:2004xg}, denoted as $F(1)$ and
$G(1)$ respectively, are shown in fig.\ref{fig:Vcb}.
%%%%%%%%%%%%%%%%%%%%%%%%%%%%%%%%%%%%%%%%%%%%%%%%%%%%%%%%%%%%%%%%%%
%%%%%%%%%%%%%%%%%%%%%%%%%%%%%%%%%%%%%%%%%%%%%%%%%%%%%%%%%%%%%%%%%%
In the infinite mass limit of both the charm and bottom quarks these form
factors are normalized to unity, up to small radiative corrections.
Fig.\ref{fig:Vcb} shows that the percent level accuracy required to evaluate the
$1/m_Q$ corrections to the static limit has been reached by the lattice
calculations. Notice that the form factors for both the $B\to D^* l \nu$ and
$B\to D l \nu$ decays have been determined by the FNAL and ROMA-TOV
collaborations using different techniques, and the agreement between the results
is reassuring. From a phenomenological point of view, the semileptonic decay in
the vector channel, $B\to D^* l \nu$, plays a privileged role, since the
experimental accuracy in the determination of the decay rate is higher by almost
a factor 10 than the one reached for $B\to D l \nu$ decays.

For the form factor $F(1)$, an unquenched determination with $N_f=2+1$ dynamical
quarks has been recently obtained by the FNAL group, and it is in good agreement
with the quenched estimates by the FNAL and ROMA-TOV collaborations (see
fig.\ref{fig:Vcb} right). Thus, we quote the unquenched result as the best
estimate of the form factor, namely
\beq
F(1)=0.924 \pm 0.022 \ .
\eeq
In the case of $B\to D l \nu$ decays, the unquenched result for $G(1)$ by the
FNAL collaboration has been only presented at the Lattice 2004
conference~\cite{Okamoto:2004xg} and it has not been submitted for publication.
Therefore, in this case, we prefer to quote as our final estimate the value
\beq
G(1)=1.060 \pm 0.035 \ ,
\eeq
obtained by combining the unquenched and the two quenched determinations by the
FNAL and ROMA-TOV collaborations, also shown in fig.\ref{fig:Vcb} (left).

In order to extract $V_{cb}$, we use the HFAG experimental averages $|V_{cb}|
\cdot F(1)=(36.18 \pm 0.55)\cdot 10^{-3}$ and $|V_{cb}|\cdot G(1) = (42.3 \pm
4.5)\cdot 10^{-3}$~\cite{hfag}, thus obtaining
\bea
\label{eq:Vcb}
&& |V_{cb}|\ {\rm (excl.)}  = (39.2 \pm 1.1)\cdot 10^{-3} 
\quad , \quad {\rm from} \ B\to D^* l \nu \nn \\
&& |V_{cb}|\ {\rm (excl.)}  = (39.9 \pm 4.4)\cdot 10^{-3} 
\quad , \quad {\rm from} \ B\to D l \nu \ .
\eea
Clearly, the determination of $V_{cb}$ is dominated by the result obtained
from $B\to D^* l \nu$ decays, which we thus take as the best estimate for the
CKM matrix element.

\section{$V_{ub}$ exclusive}
\begin{figure}[t]
\begin{center}
\includegraphics[scale=0.30,angle=270]{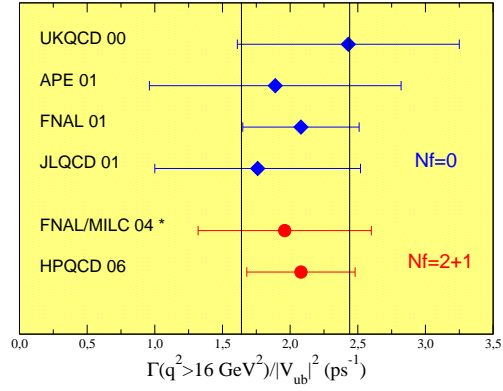}
\end{center}
\vspace{-0.5cm}
\caption{{\sl Lattice QCD results for $\Gamma(B \to \pi l \nu;\, q^2>16\
\gev^2)/|V_{ub}|^2$~\cite{Okamoto:2004xg},
\cite{Bowler:1999xn}-\cite{Dalgic:2006dt}. A star in the legend labels
preliminary results. Our average is shown by the vertical band.}}
\label{fig:Vub}
\end{figure}

Lattice QCD results for the form factor controlling the exclusive semileptonic
$B \to \pi l \nu$ decay are usually considered to evaluate the integrated decay
rate in the large momentum transfer region, which is the one directly accessible
to lattice calculations. In fig.\ref{fig:Vub} we present a compilation of both
quenched and unquenched lattice results~\cite{Okamoto:2004xg},
\cite{Bowler:1999xn}-\cite{Dalgic:2006dt} for the quantity $FF(q^2 > 16\ \gev^2)
\equiv \Gamma(q^2>16\ \gev^2)/|V_{ub}|^2$. They have been obtained using
different approaches to treat the $b$ quark (relativistic QCD, NRQCD and FNAL).
A very good agreement among all the results is observed, and the effect of the
quenching error is not visible so far within the present uncertainties.
%%%%%%%%%%%%%%%%%%%%%%%%%%%%%%%%%%%%%%%%%%%%%%%%%%%%%%%%%%%%%%%%%%
%%%%%%%%%%%%%%%%%%%%%%%%%%%%%%%%%%%%%%%%%%%%%%%%%%%%%%%%%%%%%%%%%%

By averaging the lattice results shown in fig.\ref{fig:Vub} we obtain 
\beq
FF(q^2 > 16\ \gev^2)= 2.04 \pm 0.40 \ {\rm ps^{-1}} \ ,
\eeq
which can be combined with the experimental branching fraction $BF(q^2 > 16\
\gev^2) = (0.38 \pm 0.03 \pm 0.03)\cdot 10^{-4}$ quoted for $B^0$ by
HFAG~\cite{hfag} and with the lifetime $\tau_{B^0} = 1.530 \pm 0.009 \ ps$ to
obtain
\beq
\label{eq:Vublat}
|V_{ub}|\ {\rm (excl.)}_{LQCD}  = (35.4 \pm 4.0)\cdot 10^{-4} \ .
\eeq

For these decays also the information coming from QCD sum rules on the small
momentum transfer region can be considered. Ref.~\cite{Ball:2004ye} provides
$FF(q^2 < 16\ \gev^2)= 5.44 \pm 1.43\ {\rm ps^{-1}}$ which, combined with the
HFAG average, $BF(q^2 < 16\ \gev^2) = (0.95 \pm 0.05 \pm 0.05)\cdot 10^{-4}$ and
the $B^0$ lifetime quoted above, gives $|V_{ub}|\ {\rm (excl.)}_{QCD-SR}  =
(34.7 \pm 4.8)\cdot 10^{-4}$. The more recent and updated analysis of
ref.~\cite{Duplancic:2008ix} quotes
\beq
\label{eq:Vubsr}
|V_{ub}|\ {\rm (excl.)}_{QCD-SR}  = (35 \pm 4 \pm 2 \pm 1)\cdot 10^{-4} \ ,
\eeq
in good agreement with the previous QCD sum rules result and the lattice
determination. We thus quote as our final average
\beq
\label{eq:Vub}
|V_{ub}|\ {\rm (excl.)}  = (35.0 \pm 4.0)\cdot 10^{-4} \ .
\eeq

\section{Matrix elements for $K^0-\bar K^0$ mixing: full basis of four-fermion
operators}
In models of physics beyond the Standard Model, the effective Hamiltonian which
describes the $K^0-\bar K^0$ mixing amplitude involves in general the complete
basis of $\Delta S=2$ four-fermion operators. A common choice for this basis is
constituted by the operators
\bea
\label{eq:fullop}
&& O_1 = \bar s^{\alpha}\gamma_\mu(1 -\gamma_5)d^{\alpha} 
      \bar s^{\beta}\gamma_\mu(1 - \gamma_5)d^{\beta}\ , \nn \\
&& O_2 = \bar s^{\alpha}(1 - \gamma_5)d^{\alpha} 
      \bar s^{\beta}(1 - \gamma_5)d^{\beta}\ , \nn \\
&& O_3 = \bar s^{\alpha}(1 - \gamma_5)d^{\beta} 
       \bar s^{\beta}(1 - \gamma_5)d^{\alpha} \ , \\
&& O_4 = \bar s^{\alpha}(1 - \gamma_5)d^{\alpha} 
      \bar s^{\beta}(1 + \gamma_5)d^{\beta}\ , \nn \\
&& O_5 = \bar s^{\alpha}(1 - \gamma_5)d^{\beta} 
      \bar s^{\beta}(1 + \gamma_5)d^{\alpha}\ , \nn
\eea
where $\alpha$ and $\beta$ are colour indices, together with the operators
$\tilde O_{1,2,3}$ obtained from $O_{1,2,3}$ with the exchange $\gamma_5 \to -
\gamma_5$. In chirally invariant renormalization schemes, the operators $\tilde
O_{i}$ have the same matrix elements of the $O_{i}$ and, for this reason, they
will not be further discussed in what follows.

Omitting terms which are of higher order in chiral perturbation theory, the
$B$-parameters are introduced using the expressions
\bea
\label{eq:fullbpar}
&& \langle\bar K^0\vert O_1(\mu)\vert K^0\rangle = 
   \frac{8}{3} M_K^2 f_K^2 \, B^{sd}_1(\mu) \ , \nn \\
&& \langle\bar K^0\vert O_2(\mu)\vert K^0\rangle = -\frac{5}{3} 
\left( \frac{M_K}{m_s(\mu)+m_d(\mu)} \right)^2 M_K^2 f_K^2 \, B^{sd}_2(\mu)\ ,
\nn \\
&& \langle\bar K^0\vert O_3(\mu)\vert K^0\rangle = \frac{1}{3} 
\left( \frac{M_K}{m_s(\mu)+m_d(\mu)} \right)^2 M_K^2 f_K^2 \, B^{sd}_3(\mu)\ ,
\\
&& \langle\bar K^0\vert O_4(\mu)\vert K^0\rangle = 2
\left( \frac{M_K}{m_s(\mu)+m_d(\mu)} \right)^2 M_K^2 f_K^2 \, B^{sd}_4(\mu)\ ,
\nn \\
&& \langle\bar K^0\vert O_5(\mu)\vert K^0\rangle = \frac{2}{3} 
\left( \frac{M_K}{m_s(\mu)+m_d(\mu)} \right)^2 M_K^2 f_K^2 \, B^{sd}_5(\mu)\ ,
\nn
\eea
where $B^{sd}_1(\mu)=B_K(\mu)$.

The matrix elements of the full operator basis for $K^0-\bar K^0$ mixing have
been computed, at present, only in three lattice
studies~\cite{Donini:1999nn,Babich:2006bh,Nakamura:2006eq}, all performed within
the quenched approximation. Refs.~\cite{Donini:1999nn} and \cite{Babich:2006bh}
use tree-level improved Clover and overlap fermions respectively, and both
implement non-perturbative renormalization with the RI-MOM method. The results
of ref.~\cite{Nakamura:2006eq} are obtained by using domain wall fermions and
renormalizing the operators with one-loop perturbation theory. They have been
presented at the Lattice 2006 conference but a final analysis has not been
published. The values of the $B$-parameters from
refs.~\cite{Donini:1999nn}-\cite{Nakamura:2006eq} are collected in
table~\ref{tab:BKpar} (top)~\footnote{We have selected from
ref.~\cite{Donini:1999nn} the results obtained at the largest values of the
lattice coupling, and from ref.~\cite{Nakamura:2006eq} those obtained with the
Iwasaki gauge action on the larger volume. In the latter case, we add to the
statistical error the uncertainty coming from the difference between the results
obtained with the Iwasaki and the Wilson plaquette gauge actions, which provides
an estimate of ${\cal O}(a^2)$ discretization effects.}.
%%%%%%%%%%%%%%%%%%%%%%%%%%%%%%%%%%%%%%%%%%%%%%%%%%%%%%%%%%%%%%%%%%
\begin{table}[t]
\begin{center}
\renewcommand{\arraystretch}{1.5}
\begin{tabular}{||c||c|c|c|c|c||} \cline{2-6}
%%%%%%%%%%%%%%%%%%%%%%%%%%%%%%%%%%%%%%%%%%%%%%%%%%%%%%%%%%%%%%%%%%%%%%%%
\multicolumn{1}{c||}{} 
 & $B^{sd}_1$ & $B^{sd}_2$ & $B^{sd}_3$ & $B^{sd}_4$ & $B^{sd}_5$ \\
\hline %%%%%%%%%%%%%%%%%%%%%%%%%%%%%%%%%%%%%%%%%%%%%%%%%%%%%%%%%%%%%%%%%
 Ref.~\cite{Donini:1999nn} & 
 0.68(21) & 0.67(7)  & 0.95(15)  & 1.00(9)  &  0.66(11) \\
 Ref.~\cite{Babich:2006bh}  & 
 0.56(6)  & 0.87(8)  & 1.41(16)  & 0.94(6)  &  0.62(8) \\
Ref.~\cite{Nakamura:2006eq} & 
 0.52(4)  & 0.54(2)  & 0.71(2)   & 0.70(1)  & 0.62(1) \\
\hline %%%%%%%%%%%%%%%%%%%%%%%%%%%%%%%%%%%%%%%%%%%%%%%%%%%%%%%%%%
\end{tabular}
\hspace{0.5cm}
\begin{tabular}{||c||c|c|c|c|c||} \cline{2-6}
%%%%%%%%%%%%%%%%%%%%%%%%%%%%%%%%%%%%%%%%%%%%%%%%%%%%%%%%%%%%%%%%%%%%%%%%
\multicolumn{1}{c||}{} 
& $R^{sd}_1$ & $R^{sd}_2$ & $R^{sd}_3$ & $R^{sd}_4$ & $R^{sd}_5$ \\
\hline %%%%%%%%%%%%%%%%%%%%%%%%%%%%%%%%%%%%%%%%%%%%%%%%%%%%%%%%%%%%%%%%%
Ref.~\cite{Donini:1999nn}   &  1  &  -7(2) &  1.9(6)  &   12(3) &  2.6(9) \\
Ref.~\cite{Babich:2006bh}   &  1  & -16(3) &  5.2(9)  &   21(3) &  4.6(9) \\
Ref.~\cite{Nakamura:2006eq} &  1  & -19(1) & 5.0(3)   &   30(3) &  8.8(7) \\
\hline %%%%%%%%%%%%%%%%%%%%%%%%%%%%%%%%%%%%%%%%%%%%%%%%%%%%%%%%%%%%%%%%%
\end{tabular}
\renewcommand{\arraystretch}{1.0}
\end{center}
\caption{\sl $B$-parameters (top) and ratios $R^{sd}_i$ (bottom) for the full
basis of four-fermions operators in $K^0-\bar K^0$ mixing. Results are given in
the RI-MOM scheme at the scale $\mu=2$ GeV.}
\label{tab:BKpar}
\end{table}
%%%%%%%%%%%%%%%%%%%%%%%%%%%%%%%%%%%%%%%%%%%%%%%%%%%%%%%%%%%%%%%%%%
In table~\ref{tab:BKpar} (bottom) we collect, instead, the corresponding values
of the ratios of the non Standard Model to the Standard Model matrix elements,
\beq
R^{sd}_i(\mu) = \frac{\langle\bar K^0\vert O_i(\mu)\vert K^0\rangle}
{\langle\bar K^0\vert O_1(\mu)\vert K^0\rangle} \quad , \quad i=2,\ldots 5 \ .
\eeq

From the results shown in table~\ref{tab:BKpar} we observe that the differences
among the three determinations of
refs.~\cite{Donini:1999nn}-\cite{Nakamura:2006eq} are typically larger than the
quoted uncertainties. Clearly, new lattice studies of these matrix elements
would be necessary in order to clarify the situation. The observation that the
discrepancies are much more pronounced in the case of the ratios $R_i$ might
suggest that some of the systematic errors cancel, at least in part, in the
determination of the $B$-parameters. For this reason, we average the results for
the $B$-parameters and quote:
\beq
B^{sd}_2 = 0.7(2) \quad , \quad
B^{sd}_3 = 1.0(4) \quad , \quad
B^{sd}_4 = 0.9(2) \quad , \quad
B^{sd}_5 = 0.6(1) \ ,
\eeq
in the RI-MOM scheme at the scale $\mu=2$ GeV. For the parameter $B^{sd}_1=B_K$
we refer instead to the average given in eq.~(\ref{eq:bkms}) which, translated
into the RI-MOM scheme at $\mu=2$ GeV, corresponds to $B^{sd}_1=B_K=0.54(5)$.

In order to derive from the average values of the $B$-parameters the ratios
$R^{sd}_i$, a determination of the strange and down quark masses is required.
We use, at $\mu=2$ GeV, $(m_s^{RI}+m_d^{RI})=135 \pm 18$ MeV, corresponding to
$(m_s^{\msb}+m_d^{\msb})=110 \pm 15$ MeV. In this way we find
\beq
R^{sd}_2 = -11 \pm 4 \quad , \quad
R^{sd}_3 = 3.1 \pm 1.5 \quad , \quad
R^{sd}_4 = 17 \pm 6 \quad , \quad
R^{sd}_5 = 3.8 \pm 1.3 \ .
\eeq

\section{Matrix elements for $B^0_{d/s}-\bar B^0_{d/s}$ mixing: full basis
of four-fermion operators}
In the case of $B$ mesons, besides the matrix element of the operator $O_1$
which enters the mixing amplitude, also the operator $O_2$ (sometimes denoted as
$O_S$) is of interest in the Standard Model, since it contributes to the
theoretical prediction of the lifetime difference $\Delta\Gamma$ of the neutral
$B$ mesons. As in the case of $K^0-\bar K^0$ mixing, in generic New Physics
models the effective Hamiltonian describing $\Delta B=2$ transitions receives
contribution from the full basis of four fermions operators. Here we define the
operators $O_1$-$O_5$ for $B^0_{d/s}-\bar B^0_{d/s}$ mixing and the
corresponding $B$-parameters as in eqs.~(\ref{eq:fullop}) and
(\ref{eq:fullbpar}), with the obvious replacements $s \to b$ and $d\to d/s$.

%%%%%%%%%%%%%%%%%%%%%%%%%%%%%%%%%%%%%%%%%%%%%%%%%%%%%%%%%%%%%%%%%%
\begin{table}[t]
\begin{center}
\renewcommand{\arraystretch}{1.5}
\begin{tabular}{||c||c|c|c|c|c||} \cline{2-6}
%%%%%%%%%%%%%%%%%%%%%%%%%%%%%%%%%%%%%%%%%%%%%%%%%%%%%%%%%%%%%%%%%%%%%%%%
\multicolumn{1}{c||}{} 
 & $B^{bs}_1$ & $B^{bs}_2$ & $B^{bs}_3$ & $B^{bs}_4$ & $B^{bs}_5$ \\
\hline %%%%%%%%%%%%%%%%%%%%%%%%%%%%%%%%%%%%%%%%%%%%%%%%%%%%%%%%%%%%%%%%%
Ref.~\cite{Becirevic:2001xt} &
 0.88(5)  & 0.84(4)  & 0.91(9)  & 1.15(6)  & 1.74(7)  \\
Ref.~\cite{Aoki:2002bh} &
 0.86(5)  &  0.86(5) &   ---    &   ---    &   ---    \\
Ref.~\cite{Dalgic:2006gp} &
 0.76(11) & 0.84(13) & 0.90(14) &   ---    &  ---     \\
\hline %%%%%%%%%%%%%%%%%%%%%%%%%%%%%%%%%%%%%%%%%%%%%%%%%%%%%%%%%%
\end{tabular}
\renewcommand{\arraystretch}{1.0}
\end{center}
\caption{\sl $B$-parameters for the full basis of four-fermions operators in
$B^0_s-\bar B^0_s$ mixing. The results are given at the scale $\mu=m_b$ in the
$\msb$ scheme of ref.~\cite{Beneke:1998sy} for $B_1$-$B_3$ and of
ref.~\cite{Buras:2000if} for $B_4$-$B_5$.}
\label{tab:BBpar}
\end{table}
%%%%%%%%%%%%%%%%%%%%%%%%%%%%%%%%%%%%%%%%%%%%%%%%%%%%%%%%%%%%%%%%%%
Lattice results for the full basis of $\Delta B=2$ four-fermion operators have
been obtained only in ref.~\cite{Becirevic:2001xt}, within the quenched
approximation. A quenched determination of $B_1$ and $B_2$ has been also
performed in ref.~\cite{Aoki:2002bh}~\footnote{Ref.~\cite{Aoki:2002bh} quotes
final results for the $B_2$ parameter only, although both the matrix elements of
the operators $O_2$ and $O_3$ have been evaluated, since they mix under
renormalization.}. Recently, the matrix elements of $O_1$-$O_3$ have been also
computed in a simulation with $N_f=2+1$ dynamical fermions~\cite{Dalgic:2006gp}.
All these results are collected for the $B_s$ system in table \ref{tab:BBpar},
with the $B$-parameters defined at the scale $\mu=2$ GeV in the $\msb$ scheme of
ref.~\cite{Beneke:1998sy} for $B_1$-$B_3$ and of ref.~\cite{Buras:2000if} for
$B_4$-$B_5$ (these are the renormalization schemes usually considered for these
quantities). Note that the larger errors quoted by ref.~\cite{Dalgic:2006gp} on
the $B$-parameters, with respect to refs.~\cite{Becirevic:2001xt,Aoki:2002bh},
are due to the fact that the $B$-parameters are extracted in
\cite{Dalgic:2006gp} from the values of the corresponding matrix elements,
rather than directly calculated from the more precise ratios of 2- and 3-point
correlation functions. On the other hand, in order to evaluate from the
$B$-parameters the full matrix elements of four-fermion operators, which are
eventually the quantities of physical interest, the results of
refs.~\cite{Becirevic:2001xt,Aoki:2002bh} for the $B$-parameters have to be
combined with the value of the decay constant $f_{Bs}$, which thus increases the
final uncertainty.

The $SU(3)$ breaking effects for the full basis of $B$-parameters have been
computed in ref.~\cite{Becirevic:2001xt}, that quotes
\beq
(B_i^{bs}/B_i^{bd})\vert_{i=1,\ldots,5} = \left\{
0.99(2), \, 1.01(2), \, 1.01(3), \, 1.01(2), \, 1.01(3) \right\} \ .
\eeq
These ratios are well compatible with unity within the errors. Thus, we choose
to quote common averages for $B_d$ and $B_s$ mixing.

The results collected in table~\ref{tab:BBpar} show a very good agreement among
the central values of the three (two) determinations of $B_2$ ($B_3$). On the
other hand, the comparison of several lattice determinations for the
$B$-parameter $B_1 = B_{B_q}$ has led us to quote in eq.~(\ref{eq:bbms}) a final
uncertainty on this quantity of about 10\%. Therefore, we choose to add the same
10\% of systematic uncertainty also to the final estimates of $B_2$ and $B_3$,
as well as to those of $B_4$ and $B_5$ for which only a single (and rather old)
lattice calculation exists. In this way, we quote
\beq
\label{eq:Bbms}
B_2^{bq} = 0.85(10) \quad , \quad
B_3^{bq} = 0.90(13) \quad , \quad
B_4^{bq} = 1.15(13) \quad , \quad
B_5^{bq} = 1.74(19) \ ,
\eeq
for both $q=d,s$. The results in eq.~(\ref{eq:Bbms}) are given at the scale
$\mu=m_b$ in the $\msb$ scheme of ref.~\cite{Beneke:1998sy} for $B_2$-$B_3$ and
of ref.~\cite{Buras:2000if} for $B_4$-$B_5$. For the parameter $B_1^{bs}=
B_{B_s}$ we refer instead to the average given in eq.~(\ref{eq:bbms}) (in the
same $\msb$ scheme of ref.~\cite{Beneke:1998sy}).

The values of the $B$-parameters in eq.~(\ref{eq:Bbms}), combined with the
average of the $B_s$-meson decay constant of eq.~(\ref{eq:fbsave}), allow to
evaluate the quantities
\beq
{\cal R}_i(m_b) = \left( \frac{M_{B_s}}{m_b(m_b)+m_s(m_b)} \right)\,
f_{B_s}\, \sqrt{B_i^{bs}(m_b)} \qquad , \qquad i=2,\ldots 5 \ .
\eeq
The squares ${\cal R}_i^2$ are proportional to the corresponding operator matrix
elements. Using $m_b^{\msb}(m_b)=4.21(8)$ GeV and $m_s^{\msb}(m_b)=87(12)$ MeV
(which corresponds to $m_s^{\msb}(2\ \gev)=105(15)$ MeV), we obtain
\beq
{\cal R}_2 = 282 \pm 34 \ \mev \, , \quad
{\cal R}_3 = 290 \pm 37 \ \mev \, , \quad
{\cal R}_4 = 328 \pm 39 \ \mev \, , \quad
{\cal R}_5 = 404 \pm 47 \ \mev \ .
\eeq
For ${\cal R}_2$ and ${\cal R}_3$ these results are well consistent with those
quoted by the unquenched calculation of ref.~\cite{Dalgic:2006gp}, namely ${\cal
R}_2=295 \pm 22$ MeV and ${\cal R}_3=305 \pm 23$ MeV.

\section{Matrix elements for $D^0-\bar D^0$ mixing}
The matrix elements of the full basis of four-fermion operators for $D^0-\bar
D^0$ mixing can be obtained as a byproduct from ref.~\cite{Becirevic:2001xt}.
This work provides quenched results for the $B$-parameters corresponding to
heavy-light meson masses equal to $M_P=1.75(9)$ GeV and $M_P=2.02(10)$ GeV
respectively. By interpolating between the two sets of results, one obtains for
the physical $D$-mesons the values $B_i^{cu}\vert_{i=1,\ldots,5} = 
\{0.85(2)$, 0.82(3), 1.07(5), 1.10(2), 1.37(3)\}, in
the RI-MOM scheme at the scale $\mu=2.8$ GeV. The $B$-parameter $B_D= B_1^{cu}$
has been also computed in ref.~\cite{Lin:2006vc}, always within the quenched
approximation and using non-perturbative RI-MOM renormalization, with domain
wall fermions. The result is affected by a large uncertainty: they obtain
$B_1^{cu} = 0.85(2)(11)$ in the $\msb$ scheme at $\mu=2.0$ GeV. Translated into
the RI-MOM scheme at the scale $\mu=2.8$ GeV, this value corresponds to
$B_1^{cu} = 0.81(2)(10)$, well consistent with the result of
ref.~\cite{Becirevic:2001xt}.

By considering that the results of ref.~\cite{Becirevic:2001xt} are obtained
from a single quenched lattice calculation, we add in the final averages a
systematic uncertainty of 10\%, as we have done in the case of $B^0-\bar B^0$
mixing. In this way we obtain
\bea
\label{eq:BDave}
B_1^{cu} = 0.85(9) \quad , \quad
& B_2^{cu} = 0.82(9) \quad , \quad
& B_3^{cu} = 1.07(12) \quad , \quad \nn \\
& B_4^{cu} = 1.10(11) \quad , \quad
& B_5^{cu} = 1.37(14) \quad ,
\eea
in the RI-MOM scheme at the scale $\mu=2.8$ GeV.


\begin{thebibliography}{9}

\bibitem{UTfit-ifae}
  M.~Bona {\it et al.}  [${\rm UT}fit$ Collaboration], these proceedings.

%\cite{Bona:2006ah}
\bibitem{UTfit1}
  M.~Bona {\it et al.}  [${\rm UT}fit$ Collaboration],
  %``The unitarity triangle fit in the standard model and hadronic  parameters
  %from lattice QCD: A reappraisal after the measurements of  Delta(m(s)) and
  %BR(B --> tau nu/tau),''
  JHEP {\bf 0610} (2006) 081
  [arXiv:hep-ph/0606167].
  %%CITATION = JHEPA,0610,081;%%

%\cite{Bona:2006sa}
\bibitem{UTfit2}
  M.~Bona {\it et al.}  [${\rm UT}fit$ Collaboration],
  %``The UTfit collaboration report on the unitarity triangle beyond the
  %standard model: Spring 2006,''
  Phys.\ Rev.\ Lett.\  {\bf 97} (2006) 151803
  [arXiv:hep-ph/0605213].
  %%CITATION = PRLTA,97,151803;%%

\bibitem{wwwutfit} The ${\rm UT}fit$ Collaboration, {\tt http://www.utfit.org}

%\cite{Juttner:2007sn}
\bibitem{Juttner:2007sn}
  A.~Juttner,
  %``Progress in kaon physics on the lattice,''
  PoS {\bf LATTICE2007} (2007) 014
  [arXiv:0711.1239 [hep-lat]].
  %%CITATION = POSCI,LATTICE2007,014;%%

%%%%%%%%%%%%%%%%%%%%%%%%%%%%%%%%%%%%%%%%%%%%%%%%%%%%%%%%%%%%%%%%%%%%%%%
%%%%%%%%%%%%%   BK
%%%%%%%%%%%%%%%%%%%%%%%%%%%%%%%%%%%%%%%%%%%%%%%%%%%%%%%%%%%%%%%%%%%%%%%

%\cite{Aoki:1997nr}
\bibitem{Aoki:1997nr}
  S.~Aoki {\it et al.}  [JLQCD Collaboration],
  %``Kaon B parameter from quenched lattice QCD,''
  Phys.\ Rev.\ Lett.\  {\bf 80} (1998) 5271
  [arXiv:hep-lat/9710073].
  %%CITATION = PRLTA,80,5271;%%

%\cite{Aoki:2005ga}
\bibitem{Aoki:2005ga}
  Y.~Aoki {\it et al.},
  %``The kaon B-parameter from quenched domain-wall QCD,''
  Phys.\ Rev.\  D {\bf 73} (2006) 094507
  [arXiv:hep-lat/0508011].
  %%CITATION = PHRVA,D73,094507;%%

%\cite{Nakamura:2008xz}
\bibitem{Nakamura:2008xz}
  Y.~Nakamura, S.~Aoki, Y.~Taniguchi and T.~Yoshie  [CP-PACS Collaboration],
  %``Precise determination of $B_K$ and right quark masses in quenched
  %domain-wall QCD,''
  arXiv:0803.2569 [hep-lat].
  %%CITATION = ARXIV:0803.2569;%%

%\cite{Aoki:2004ht}
\bibitem{Aoki:2004ht}
  Y.~Aoki {\it et al.},
  %``Lattice QCD with two dynamical flavors of domain wall fermions,''
  Phys.\ Rev.\  D {\bf 72} (2005) 114505
  [arXiv:hep-lat/0411006].
  %%CITATION = PHRVA,D72,114505;%%

%\cite{Aoki:2008ss}
\bibitem{Aoki:2008ss}
  S.~Aoki {\it et al.}  [JLQCD Collaboration],
  %``B_K with two flavors of dynamical overlap fermions,''
  arXiv:0801.4186 [hep-lat].
  %%CITATION = ARXIV:0801.4186;%%

%\cite{Gamiz:2006sq}
\bibitem{Gamiz:2006sq}
  E.~Gamiz, S.~Collins, C.~T.~H.~Davies, G.~P.~Lepage, J.~Shigemitsu and
  M.~Wingate [HPQCD Collaboration],
  %``Unquenched determination of the kaon parameter B(K) from improved
  %staggered fermions,''
  Phys.\ Rev.\  D {\bf 73} (2006) 114502
  [arXiv:hep-lat/0603023].
  %%CITATION = PHRVA,D73,114502;%%

%\cite{Antonio:2007pb}
\bibitem{Antonio:2007pb}
  D.~J.~Antonio {\it et al.}  [RBC Collaboration],
  %``Neutral kaon mixing from 2+1 flavor domain wall QCD,''
  Phys.\ Rev.\ Lett.\  {\bf 100} (2008) 032001
  [arXiv:hep-ph/0702042].
  %%CITATION = PRLTA,100,032001;%%

%%%%%%%%%%%%%%%%%%%%%%%%%%%%%%%%%%%%%%%%%%%%%%%%%%%%%%%%%%%%%%%%%%%%%%%

%\cite{Dawson:2005za}
\bibitem{dawson}
  C.~Dawson,
  %``Progress In Kaon Phenomenology From Lattice QCD,''
  PoS {\bf LAT2005} (2006) 007.
  %%CITATION = POSCI,LAT2005,007;%%

%%%%%%%%%%%%%%%%%%%%%%%%%%%%%%%%%%%%%%%%%%%%%%%%%%%%%%%%%%%%%%%%%%%%%%%
%%%%%%%%%%%%%   fB,fBs
%%%%%%%%%%%%%%%%%%%%%%%%%%%%%%%%%%%%%%%%%%%%%%%%%%%%%%%%%%%%%%%%%%%%%%%

%\cite{AliKhan:2000eg}
\bibitem{AliKhan:2000eg}
  A.~Ali Khan {\it et al.}  [CP-PACS Collaboration],
  %``Decay constants of B and D mesons from improved relativistic lattice  QCD
  %with two flavours of sea quarks,''
  Phys.\ Rev.\  D {\bf 64} (2001) 034505
  [arXiv:hep-lat/0010009].
  %%CITATION = PHRVA,D64,034505;%%

%\cite{AliKhan:2001jg}
\bibitem{AliKhan:2001jg}
  A.~Ali Khan {\it et al.}  [CP-PACS Collaboration],
  %``B meson decay constant from two-flavor lattice QCD with  non-relativistic
  %heavy quarks,''
  Phys.\ Rev.\  D {\bf 64} (2001) 054504
  [arXiv:hep-lat/0103020].
  %%CITATION = PHRVA,D64,054504;%%

%\cite{Bernard:2002pc}
\bibitem{Bernard:2002pc}
  C.~Bernard {\it et al.}  [MILC Collaboration],
  %``Lattice calculation of heavy-light decay constants with two flavors of
  %dynamical quarks,''
  Phys.\ Rev.\  D {\bf 66} (2002) 094501
  [arXiv:hep-lat/0206016].
  %%CITATION = PHRVA,D66,094501;%%

%\cite{Aoki:2003xb}
\bibitem{Aoki:2003xb}
  S.~Aoki {\it et al.}  [JLQCD Collaboration],
  %``B0 anti-B0 mixing in unquenched lattice QCD,''
  Phys.\ Rev.\ Lett.\  {\bf 91} (2003) 212001
  [arXiv:hep-ph/0307039].
  %%CITATION = PRLTA,91,212001;%%

%\cite{Wingate:2003gm}
\bibitem{Wingate:2003gm}
  M.~Wingate, C.~T.~H.~Davies, A.~Gray, G.~P.~Lepage and J.~Shigemitsu,
  %``The B/s and D/s decay constants in 3 flavor lattice QCD,''
  Phys.\ Rev.\ Lett.\  {\bf 92} (2004) 162001
  [arXiv:hep-ph/0311130].
  %%CITATION = PRLTA,92,162001;%%

%\cite{Gray:2005ad}
\bibitem{Gray:2005ad}
  A.~Gray {\it et al.}  [HPQCD Collaboration],
  %``The B meson decay constant from unquenched lattice QCD,''
  Phys.\ Rev.\ Lett.\  {\bf 95} (2005) 212001
  [arXiv:hep-lat/0507015].
  %%CITATION = PRLTA,95,212001;%%

%\cite{Bernard:2007}
\bibitem{Bernard:2007}
  C. Bernard {\it et al.} [Fermilab Lattice and MILC Collaborations],
  %``The decay constants fB and fD+ from three-flavor lattice QCD''
  PoS {\bf LATTICE2007} (2007) 370
  %%CITATION = POSCI,LATTICE2007,370;%%
%%%%%%%%%%%%%%%%%%%%%%%%%%%%%%%%%%%%%%%%%%%%%%%%%%%%%%%%%%%%%%%%%%%%%%%

%\cite{ElKhadra:1996mp}
\bibitem{ElKhadra:1996mp}
  A.~X.~El-Khadra, A.~S.~Kronfeld and P.~B.~Mackenzie,
  %``Massive Fermions in Lattice Gauge Theory,''
  Phys.\ Rev.\  D {\bf 55} (1997) 3933
  [arXiv:hep-lat/9604004].
  %%CITATION = PHRVA,D55,3933;%%

%%%%%%%%%%%%%%%%%%%%%%%%%%%%%%%%%%%%%%%%%%%%%%%%%%%%%%%%%%%%%%%%%%%%%%%
%%%%%%%%%%%%%   fD,fDs
%%%%%%%%%%%%%%%%%%%%%%%%%%%%%%%%%%%%%%%%%%%%%%%%%%%%%%%%%%%%%%%%%%%%%%%

%\cite{Rosner:2008yu}
\bibitem{Rosner:2008yu}
  J.~L.~Rosner and S.~Stone,
  %``Decay Constants of Charged Pseudoscalar Mesons,''
  arXiv:0802.1043 [hep-ex].
  %%CITATION = ARXIV:0802.1043;%%

%%%%%%%%%%%%%%%%%%%%%%%%%%%%%%%%%%%%%%%%%%%%%%%%%%%%%%%%%%%%%%%%%%%%%%%

% %\cite{AliKhan:2000eg}
% \bibitem{AliKhan:2000eg}
%   A.~Ali Khan {\it et al.}  [CP-PACS Collaboration],
%   %``Decay constants of B and D mesons from improved relativistic lattice  QCD
%   %with two flavours of sea quarks,''
%   Phys.\ Rev.\  D {\bf 64} (2001) 034505
%   [arXiv:hep-lat/0010009].
%   %%CITATION = PHRVA,D64,034505;%%

% %\cite{Bernard:2002pc}
% \bibitem{Bernard:2002pc}
%   C.~Bernard {\it et al.}  [MILC Collaboration],
%   %``Lattice calculation of heavy-light decay constants with two flavors of
%   %dynamical quarks,''
%   Phys.\ Rev.\  D {\bf 66} (2002) 094501
%   [arXiv:hep-lat/0206016].
%   %%CITATION = PHRVA,D66,094501;%%

\bibitem{cecilia_lat08}
  C.~Tarantino [ETM Collaboration], talk at Lattice'08, \\
  {\tt http://conferences.jlab.org/lattice2008}

% %\cite{Bernard:2007}
% \bibitem{Bernard:2007}
%   C. Bernard {\it et al.} [Fermilab Lattice and MILC Collaborations],
%   %``The decay constants fB and fD+ from three-flavor lattice QCD''
%   PoS {\bf LATTICE2007} (2007) 370
%   %%CITATION = POSCI,LATTICE2007,370;%%

%\cite{Follana:2007uv}
\bibitem{Follana:2007uv}
  E.~Follana, C.~T.~H.~Davies, G.~P.~Lepage and J.~Shigemitsu  [HPQCD
                  Collaboration and UKQCD Collaboration],
  %``High Precision determination of the pi, K, D and D_s decay constants   from
  %lattice QCD,''
  Phys.\ Rev.\ Lett.\  {\bf 100} (2008) 062002
  [arXiv:0706.1726 [hep-lat]].
  %%CITATION = PRLTA,100,062002;%%

%%%%%%%%%%%%%%%%%%%%%%%%%%%%%%%%%%%%%%%%%%%%%%%%%%%%%%%%%%%%%%%%%%%%%%%
%\cite{Dobrescu:2008er}
\bibitem{Dobrescu:2008er}
  B.~A.~Dobrescu and A.~S.~Kronfeld,
  %``Accumulating evidence for nonstandard leptonic decays of D_s mesons,''
  Phys.\ Rev.\ Lett.\  {\bf 100} (2008) 241802
  [arXiv:0803.0512 [hep-ph]].
  %%CITATION = PRLTA,100,241802;%%

%\cite{Sharpe:2006re}
\bibitem{Sharpe:2006re}
  S.~R.~Sharpe,
  %``Rooted staggered fermions: Good, bad or ugly?,''
  PoS {\bf LAT2006} (2006) 022
  [arXiv:hep-lat/0610094].
  %%CITATION = POSCI,LAT2006,022;%%

%\cite{Kronfeld:2007ek}
\bibitem{Kronfeld:2007ek}
  A.~S.~Kronfeld,
  %``Lattice gauge theory with staggered fermions: how, where, and why (not),''
  PoS {\bf LATTICE2007} (2007) 016
  [arXiv:0711.0699 [hep-lat]].
  %%CITATION = POSCI,LATTICE2007,016;%%

%%%%%%%%%%%%%%%%%%%%%%%%%%%%%%%%%%%%%%%%%%%%%%%%%%%%%%%%%%%%%%%%%%%%%%%
%%%%%%%%%%%%%   BB,BBs
%%%%%%%%%%%%%%%%%%%%%%%%%%%%%%%%%%%%%%%%%%%%%%%%%%%%%%%%%%%%%%%%%%%%%%%

%\cite{Lellouch:2000tw}
\bibitem{Lellouch:2000tw}
  L.~Lellouch and C.~J.~D.~Lin  [UKQCD Collaboration],
  %``Standard model matrix elements for neutral B meson mixing and  associated
  %decay constants,''
  Phys.\ Rev.\  D {\bf 64} (2001) 094501
  [arXiv:hep-ph/0011086].
  %%CITATION = PHRVA,D64,094501;%%

%\cite{Becirevic:2001xt}
\bibitem{Becirevic:2001xt}
  D.~Becirevic, V.~Gimenez, G.~Martinelli, M.~Papinutto and J.~Reyes,
  %``B-parameters of the complete set of matrix elements of Delta(B) = 2
  %operators from the lattice,''
  JHEP {\bf 0204} (2002) 025
  [arXiv:hep-lat/0110091].
  %%CITATION = JHEPA,0204,025;%%

%\cite{Aoki:2002bh}
\bibitem{Aoki:2002bh}
  S.~Aoki {\it et al.}  [JLQCD Collaboration],
  %``B0 - anti-B0 mixing in quenched lattice QCD. ((U)) ((W)),''
  Phys.\ Rev.\  D {\bf 67} (2003) 014506
  [arXiv:hep-lat/0208038].
  %%CITATION = PHRVA,D67,014506;%%

% %\cite{Aoki:2003xb}
% \bibitem{Aoki:2003xb}
%   S.~Aoki {\it et al.}  [JLQCD Collaboration],
%   %``B0 anti-B0 mixing in unquenched lattice QCD,''
%   Phys.\ Rev.\ Lett.\  {\bf 91} (2003) 212001
%   [arXiv:hep-ph/0307039].
%   %%CITATION = PRLTA,91,212001;%%

%\cite{Dalgic:2006gp}
\bibitem{Dalgic:2006gp}
  E.~Dalgic {\it et al.},
  %``B/s0 - anti-B/s0 mixing parameters from unquenched lattice QCD,''
  Phys.\ Rev.\  D {\bf 76} (2007) 011501
  [arXiv:hep-lat/0610104].
  %%CITATION = PHRVA,D76,011501;%%

%\cite{Albertus}
\bibitem{Albertus}
  C. Albertus {\it et al.}  [RBC and UKQCD Collaborations],
  %``B−B-Mixing with Domain Wall Fermions in the Static Approximation''
  PoS {\bf LAT2007} (2007) 376
  %%CITATION = POSCI,LAT2007,376;%%

%%%%%%%%%%%%%%%%%%%%%%%%%%%%%%%%%%%%%%%%%%%%%%%%%%%%%%%%%%%%%%%%%%%%%%%
%%%%%%%%%%%%%   Vcb
%%%%%%%%%%%%%%%%%%%%%%%%%%%%%%%%%%%%%%%%%%%%%%%%%%%%%%%%%%%%%%%%%%%%%%%

%\cite{Hashimoto:2001nb}
\bibitem{Hashimoto:2001nb}
  S.~Hashimoto, A.~S.~Kronfeld, P.~B.~Mackenzie, S.~M.~Ryan and J.~N.~Simone,
  %``Lattice calculation of the zero recoil form factor of anti-B --> D* l
  %anti-nu: Toward a model independent determination of |V(cb)|,''
  Phys.\ Rev.\  D {\bf 66} (2002) 014503
  [arXiv:hep-ph/0110253].
  %%CITATION = PHRVA,D66,014503;%%

%\cite{deDivitiis:2008df}
\bibitem{deDivitiis:2008df}
  G.~M.~de Divitiis, R.~Petronzio and N.~Tantalo,
  %``Quenched lattice calculation of the vector channel B --> D* l nu decay
  %rate,''
  arXiv:0807.2944 [hep-lat].
  %%CITATION = ARXIV:0807.2944;%%

%\cite{Laiho:2007pn}
\bibitem{Laiho:2007pn}
  J.~Laiho  [Fermilab Lattice and MILC Collaborations],
  %``B -> D* l nu with 2+1 flavors,''
  PoS {\bf LATTICE2007} (2006) 358
  [arXiv:0710.1111 [hep-lat]].
  %%CITATION = POSCI,LATTICE2007,358;%%

%\cite{Hashimoto:1999yp}
\bibitem{Hashimoto:1999yp}
  S.~Hashimoto, A.~X.~El-Khadra, A.~S.~Kronfeld, P.~B.~Mackenzie, S.~M.~Ryan and
  J.~N.~Simone,
  %``Lattice {QCD} calculation of anti-B --> D l anti-nu decay form factors  at
  %zero recoil,''
  Phys.\ Rev.\  D {\bf 61} (2000) 014502
  [arXiv:hep-ph/9906376].
  %%CITATION = PHRVA,D61,014502;%%

%\cite{de Divitiis:2007ui}
\bibitem{de Divitiis:2007ui}
  G.~M.~de Divitiis, E.~Molinaro, R.~Petronzio and N.~Tantalo,
  %``Quenched lattice calculation of the B --> D l nu decay rate,''
  Phys.\ Lett.\  B {\bf 655} (2007) 45
  [arXiv:0707.0582 [hep-lat]].
  %%CITATION = PHLTA,B655,45;%%

%\cite{Okamoto:2004xg}
\bibitem{Okamoto:2004xg}
  M.~Okamoto {\it et al.},
  %``Semileptonic D --> pi / K and B --> pi / D decays in 2+1 flavor lattice
  %QCD,''
  Nucl.\ Phys.\ Proc.\ Suppl.\  {\bf 140} (2005) 461
  [arXiv:hep-lat/0409116].
  %%CITATION = NUPHZ,140,461;%%

%%%%%%%%%%%%%%%%%%%%%%%%%%%%%%%%%%%%%%%%%%%%%%%%%%%%%%%%%%%%%%%%%%%%%%%

\bibitem{hfag} The Heavy Flavor Averaging Group (HFAG),
{\tt http://www.slac.stanford.edu/xorg/hfag}

%%%%%%%%%%%%%%%%%%%%%%%%%%%%%%%%%%%%%%%%%%%%%%%%%%%%%%%%%%%%%%%%%%%%%%%
%%%%%%%%%%%%%   Vub
%%%%%%%%%%%%%%%%%%%%%%%%%%%%%%%%%%%%%%%%%%%%%%%%%%%%%%%%%%%%%%%%%%%%%%%

%\cite{Bowler:1999xn}
\bibitem{Bowler:1999xn}
  K.~C.~Bowler {\it et al.}  [UKQCD Collaboration],
  %``Improved B --> pi l nu/l form factors from the lattice,''
  Phys.\ Lett.\  B {\bf 486} (2000) 111
  [arXiv:hep-lat/9911011].
  %%CITATION = PHLTA,B486,111;%%

%\cite{Abada:2000ty}
\bibitem{Abada:2000ty}
  A.~Abada, D.~Becirevic, P.~Boucaud, J.~P.~Leroy, V.~Lubicz and F.~Mescia,
  %``Heavy --> light semileptonic decays of pseudoscalar mesons from lattice
  %QCD,''
  Nucl.\ Phys.\  B {\bf 619} (2001) 565
  [arXiv:hep-lat/0011065].
  %%CITATION = NUPHA,B619,565;%%

%\cite{ElKhadra:2001rv}
\bibitem{ElKhadra:2001rv}
  A.~X.~El-Khadra, A.~S.~Kronfeld, P.~B.~Mackenzie, S.~M.~Ryan and J.~N.~Simone,
  %``The semileptonic decays B --> pi l nu and D --> pi l nu from lattice
  %QCD,''
  Phys.\ Rev.\  D {\bf 64} (2001) 014502
  [arXiv:hep-ph/0101023].
  %%CITATION = PHRVA,D64,014502;%%

%\cite{Aoki:2001rd}
\bibitem{Aoki:2001rd}
  S.~Aoki {\it et al.}  [JLQCD Collaboration],
  %``Differential decay rate of B --> pi l nu semileptonic decay with  lattice
  %NRQCD,''
  Phys.\ Rev.\  D {\bf 64} (2001) 114505
  [arXiv:hep-lat/0106024].
  %%CITATION = PHRVA,D64,114505;%%

% %\cite{Okamoto:2004xg}
% \bibitem{Okamoto:2004xg}
%   M.~Okamoto {\it et al.},
%   %``Semileptonic D --> pi / K and B --> pi / D decays in 2+1 flavor lattice
%   %QCD,''
%   Nucl.\ Phys.\ Proc.\ Suppl.\  {\bf 140} (2005) 461
%   [arXiv:hep-lat/0409116].
%   %%CITATION = NUPHZ,140,461;%%

%\cite{Dalgic:2006dt}
\bibitem{Dalgic:2006dt}
  E.~Dalgic, A.~Gray, M.~Wingate, C.~T.~H.~Davies, G.~P.~Lepage and
  J.~Shigemitsu,
  %``B Meson Semileptonic Form Factors from Unquenched Lattice QCD,''
  Phys.\ Rev.\  D {\bf 73} (2006) 074502
  [Erratum-ibid.\  D {\bf 75} (2007) 119906]
  [arXiv:hep-lat/0601021].
  %%CITATION = PHRVA,D73,074502;%%

%\cite{Ball:2004ye}
\bibitem{Ball:2004ye}
  P.~Ball and R.~Zwicky,
  %``New results on B --> pi, K, eta decay formfactors from light-cone sum
  %rules,''
  Phys.\ Rev.\  D {\bf 71} (2005) 014015
  [arXiv:hep-ph/0406232].
  %%CITATION = PHRVA,D71,014015;%%

%\cite{Duplancic:2008ix}
\bibitem{Duplancic:2008ix}
  G.~Duplancic, A.~Khodjamirian, T.~Mannel, B.~Melic and N.~Offen,
  %``Light-cone sum rules for $B \to \pi$ form factors revisited,''
  JHEP {\bf 0804} (2008) 014
  [arXiv:0801.1796 [hep-ph]].
  %%CITATION = JHEPA,0804,014;%%

%%%%%%%%%%%%%%%%%%%%%%%%%%%%%%%%%%%%%%%%%%%%%%%%%%%%%%%%%%%%%%%%%%%%%%%
%%%%%%%%%%%%%   K-Kbar complete basis
%%%%%%%%%%%%%%%%%%%%%%%%%%%%%%%%%%%%%%%%%%%%%%%%%%%%%%%%%%%%%%%%%%%%%%%

%\cite{Donini:1999nn}
\bibitem{Donini:1999nn}
  A.~Donini, V.~Gimenez, L.~Giusti and G.~Martinelli,
  %``Renormalization group invariant matrix elements of Delta(S) = 2 and
  %Delta(I) = 3/2 four-fermion operators without quark masses,''
  Phys.\ Lett.\  B {\bf 470} (1999) 233
  [arXiv:hep-lat/9910017].
  %%CITATION = PHLTA,B470,233;%%

%\cite{Babich:2006bh}
\bibitem{Babich:2006bh}
  R.~Babich, N.~Garron, C.~Hoelbling, J.~Howard, L.~Lellouch and C.~Rebbi,
  %``K0 anti-K0 mixing beyond the standard model and CP-violating  electroweak
  %penguins in quenched QCD with exact chiral symmetry,''
  Phys.\ Rev.\  D {\bf 74} (2006) 073009
  [arXiv:hep-lat/0605016].
  %%CITATION = PHRVA,D74,073009;%%

%\cite{Nakamura:2006eq}
\bibitem{Nakamura:2006eq}
  Y.~Nakamura {\it et al.}  [CP-PACS Collaboration],
  %``Kaon B-parameters for generic Delta(S) = 2 four-quark operators in quenched
  %domain wall QCD,''
  PoS {\bf LAT2006} (2006) 089
  [arXiv:hep-lat/0610075].
  %%CITATION = POSCI,LAT2006,089;%%

%%%%%%%%%%%%%%%%%%%%%%%%%%%%%%%%%%%%%%%%%%%%%%%%%%%%%%%%%%%%%%%%%%%%%%%
%%%%%%%%%%%%%   B-Bbar complete (or almost) basis
%%%%%%%%%%%%%%%%%%%%%%%%%%%%%%%%%%%%%%%%%%%%%%%%%%%%%%%%%%%%%%%%%%%%%%%

% %\cite{Becirevic:2001xt}
% \bibitem{Becirevic:2001xt}
%   D.~Becirevic, V.~Gimenez, G.~Martinelli, M.~Papinutto and J.~Reyes,
%   %``B-parameters of the complete set of matrix elements of Delta(B) = 2
%   %operators from the lattice,''
%   JHEP {\bf 0204} (2002) 025
%   [arXiv:hep-lat/0110091].
%   %%CITATION = JHEPA,0204,025;%%

% %\cite{Aoki:2002bh}
% \bibitem{Aoki:2002bh}
%   S.~Aoki {\it et al.}  [JLQCD Collaboration],
%   %``B0 - anti-B0 mixing in quenched lattice QCD. ((U)) ((W)),''
%   Phys.\ Rev.\  D {\bf 67} (2003) 014506
%   [arXiv:hep-lat/0208038].
%   %%CITATION = PHRVA,D67,014506;%%

% %\cite{Dalgic:2006gp}
% \bibitem{Dalgic:2006gp}
%   E.~Dalgic {\it et al.},
%   %``B/s0 - anti-B/s0 mixing parameters from unquenched lattice QCD,''
%   Phys.\ Rev.\  D {\bf 76} (2007) 011501
%   [arXiv:hep-lat/0610104].
%   %%CITATION = PHRVA,D76,011501;%%

%%%%%%%%%%%%%%%%%%%%%%%%%%%%%%%%%%%%%%%%%%%%%%%%%%%%%%%%%%%%%%%%%%%%%%%

%\cite{Beneke:1998sy}
\bibitem{Beneke:1998sy}
  M.~Beneke, G.~Buchalla, C.~Greub, A.~Lenz and U.~Nierste,
  %``Next-to-leading order {QCD} corrections to the lifetime difference of  B/s
  %mesons,''
  Phys.\ Lett.\  B {\bf 459} (1999) 631
  [arXiv:hep-ph/9808385].
  %%CITATION = PHLTA,B459,631;%%

%\cite{Buras:2000if}
\bibitem{Buras:2000if}
  A.~J.~Buras, M.~Misiak and J.~Urban,
  %``Two-loop QCD anomalous dimensions of flavour-changing four-quark  operators
  %within and beyond the standard model,''
  Nucl.\ Phys.\  B {\bf 586} (2000) 397
  [arXiv:hep-ph/0005183].
  %%CITATION = NUPHA,B586,397;%%

%%%%%%%%%%%%%%%%%%%%%%%%%%%%%%%%%%%%%%%%%%%%%%%%%%%%%%%%%%%%%%%%%%%%%%%
%%%%%%%%%%%%%   D-Dbar
%%%%%%%%%%%%%%%%%%%%%%%%%%%%%%%%%%%%%%%%%%%%%%%%%%%%%%%%%%%%%%%%%%%%%%%

% %\cite{Becirevic:2001xt}
% \bibitem{Becirevic:2001xt}
%   D.~Becirevic, V.~Gimenez, G.~Martinelli, M.~Papinutto and J.~Reyes,
%   %``B-parameters of the complete set of matrix elements of Delta(B) = 2
%   %operators from the lattice,''
%   JHEP {\bf 0204} (2002) 025
%   [arXiv:hep-lat/0110091].
%   %%CITATION = JHEPA,0204,025;%%

%\cite{Lin:2006vc}
\bibitem{Lin:2006vc}
  H.~W.~Lin, S.~Ohta, A.~Soni and N.~Yamada,
  %``Charm as a domain wall fermion in quenched lattice QCD,''
  Phys.\ Rev.\  D {\bf 74} (2006) 114506
  [arXiv:hep-lat/0607035].
  %%CITATION = PHRVA,D74,114506;%%

\end{thebibliography}
\end{document}